\pdfoutput=1
\documentclass[twocolumn]{aastex631}

\newcommand{\arhp}{ArH$^+$} 
\newcommand{\ohp}{OH$^+$} 
\newcommand{\htop}{H$_{2}$O$^+$} 
\newcommand{\cii}{C\,{\small II}} 
\newcommand{\oi}{O\,{\small I}} 
\newcommand{\hi}{H\,{\small I}} 
\usepackage{comment}
\usepackage{amsmath}
\usepackage{enumitem}
\usepackage{float}
\usepackage{inputenc}
\graphicspath{{./}{Figures/}}
\shorttitle{HyGAL: Characterizing the Galactic ISM with hydrides}
\shortauthors{Jacob et al.}

\begin{document}

\title{HyGAL: Characterizing the Galactic ISM with observations of hydrides and other small molecules – I.  Survey description and a first look toward W3(OH), W3~IRS5 and NGC~7538~IRS1}

\correspondingauthor{Arshia M. Jacob}
\email{ajacob51@jhu.edu}

\author[0000-0001-7838-3425
]{A. M. Jacob}
\affiliation{William H. Miller III Department of Physics \& Astronomy, Johns Hopkins University, Baltimore, MD 21218, USA }
\affiliation{Max-Planck-Institut f\"{u}r Radioastronomie, Auf dem H\"{u}gel 69, 53121 Bonn, Germany}
\author[0000-0001-8341-1646]{D. A. Neufeld}
\affiliation{William H. Miller III Department of Physics \& Astronomy, Johns Hopkins University, Baltimore, MD 21218, USA }
\author[0000-0003-2141-5689]{P. Schilke}
\affiliation{I. Physikalisches Institut, Universit\"{a}t zu K\"{o}ln, Z\"{u}lpicher Str. 77, 50937 K\"{o}ln, Germany}
\author[0000-0002-5135-8657]{H. Wiesemeyer}
\affiliation{Max-Planck-Institut f\"{u}r Radioastronomie, Auf dem H\"{u}gel 69, 53121 Bonn, Germany}
\author[0000-0003-0364-6715]{W. Kim}
\affiliation{I. Physikalisches Institut, Universit\"{a}t zu K\"{o}ln, Z\"{u}lpicher Str. 77, 50937 K\"{o}ln, Germany}
\author[0000-0002-0404-003X]{S. Bialy}
\affiliation{Department of Astronomy, University of Maryland, College Park, MD 20742-2421, USA}
\author[0000-0003-4961-6511]{M. Busch}
\affiliation{William H. Miller III Department of Physics \& Astronomy, Johns Hopkins University, Baltimore, MD 21218, USA }
\author[0000-0002-9120-5890]{D. Elia}
\affiliation{INAF - Istituto di Astrofisica e Planetologia Spaziali, via Fosso del Cavaliere 100, 00133, Roma, Italy}
\author[0000-0003-0693-2477]{E. Falgarone}
\affiliation{Laboratoire de Physique de l'ENS, ENS, Universit\'{e} PSL, CNRS, Sorbonne Universit\'{e}, Universit\'{e} de Paris, 24 rue Lhomond, 75005 Paris, France}
\author[0000-0002-2418-7952]{M. Gerin}
\affiliation{LERMA, Observatoire de Paris, PSL Research University, CNRS, Sorbonne Universités, F-75014 Paris, France}
\author{B. Godard}
\affiliation{Observatoire de Paris, \'{E}cole normale sup\'{e}rieure, Universit\'{e} PSL, Sorbonne Universit\'{e}, CNRS, LERMA, 75005, Paris, France}
\author[0000-0001-8195-3900]{R. Higgins}
\affiliation{I. Physikalisches Institut, Universit\"{a}t zu K\"{o}ln, Z\"{u}lpicher Str. 77, 50937 K\"{o}ln, Germany}
\author[0000-0002-0472-7202]{P. Hennebelle}
\affiliation{AIM, CEA, CNRS, Universit\'{e} Paris-Saclay, Universit\'{e} Paris Diderot, Sorbonne Paris Cit\'{e}, 91191, Gif-sur-Yvette, France}
\author[0000-0001-8533-6440]{N. Indriolo}
\affiliation{AURA for the European Space Agency (ESA), Space Telescope Science Institute, 3700 San Martin Drive, Baltimore, MD 21218, USA}
\author[0000-0002-0500-4700]{D. C. Lis}
\affiliation{Jet Propulsion Laboratory, California Institute of Technology, 4800 Oak Grove Drive, Pasadena, CA, 91109, USA}
\author[0000-0001-6459-0669]{K. M. Menten}
\affiliation{Max-Planck-Institut f\"{u}r Radioastronomie, Auf dem H\"{u}gel 69, 53121 Bonn, Germany}
\author[0000-0002-3078-9482]{A. Sanchez-Monge}
\affiliation{I. Physikalisches Institut, Universit\"{a}t zu K\"{o}ln, Z\"{u}lpicher Str. 77, 50937 K\"{o}ln, Germany}
\author[0000-0002-8351-3877]{V. Ossenkopf-Okada}
\affiliation{I. Physikalisches Institut, Universit\"{a}t zu K\"{o}ln, Z\"{u}lpicher Str. 77, 50937 K\"{o}ln, Germany}
\author{M. R. Rugel}
\affiliation{Max-Planck-Institut f\"{u}r Radioastronomie, Auf dem H\"{u}gel 69, 53121 Bonn, Germany}
\author[0000-0002-0368-9160]{D. Seifried}
\affiliation{I. Physikalisches Institut, Universit\"{a}t zu K\"{o}ln, Z\"{u}lpicher Str. 77, 50937 K\"{o}ln, Germany}
\author[0000-0003-2027-5020]{P. Sonnentrucker}
\affiliation{European Space Agency (ESA), ESA Office, Space Telescope Science Institute, 3700 San Martin Drive, Baltimore, MD 21218, USA}
\author[0000-0001-6941-7638]{S. Walch}
\affiliation{I. Physikalisches Institut, Universit\"{a}t zu K\"{o}ln, Z\"{u}lpicher Str. 77, 50937 K\"{o}ln, Germany}
\author[0000-0003-0030-9510]{M. Wolfire}
\affiliation{Department of Astronomy, University of Maryland, College Park, MD 20742-2421, USA}
\author[0000-0003-4516-3981]{F. Wyrowski}
\affiliation{Max-Planck-Institut f\"{u}r Radioastronomie, Auf dem H\"{u}gel 69, 53121 Bonn, Germany}
\author[0000-0002-5595-822X]{V. Valdivia}
\affiliation{Laboratoire AIM, Paris-Saclay, CEA/IRFU/SAp-CNRS-Universit\'{e} Paris Diderot, F-91191 Gif-sur-Yvette Cedex, France}
\affiliation{Department of Physics, Graduate School of Science, Nagoya University, Furo-cho, Chikusa-ku, Nagoya 464-8602, Japan }

\begin{abstract}

The HyGAL SOFIA legacy program surveys six hydride molecules -- ArH$^+$, OH$^+$, H$_2$O$^+$, SH, OH, and CH -- and two atomic constituents -- C$^+$ and O -- within the diffuse interstellar medium (ISM) by means of absorption-line spectroscopy toward 25 bright Galactic background continuum sources. This detailed spectroscopic study is designed to exploit the unique value of specific hydrides as tracers and probes of different phases of the ISM, as demonstrated by recent studies with the \textit{Herschel} Space Observatory. 
The observations performed under the HyGAL program will allow us to address several questions related to the lifecycle of molecular material in the ISM and the physical processes that impact the phase transition from atomic to molecular gas, such as: (1) What is the distribution function of the H$_2$ fraction in the ISM? (2) How does the ionization rate due to low-energy cosmic-rays vary within the Galaxy? (3) What is the nature of interstellar turbulence (e.g.,\ typical shear or shock velocities), and what mechanisms lead to its dissipation? In this overview, we discuss the observing strategy, the synergies with ancillary and archival observations of other small molecules, and the data reduction and analysis schemes we adopted; and we present the first results obtained toward three of the survey targets, W3(OH), W3~IRS5 and NGC~7538~IRS1. 
Robust measurements of the column densities of these hydrides -- obtained through widespread observations of absorption lines-- help address the questions raised, and there is a very timely synergy between these observations and the development of theoretical models, particularly pertaining to the formation of H$_{2}$ within the turbulent ISM. The provision of enhanced HyGAL data products will therefore serve as a legacy for future ISM studies.

\end{abstract}


\section{Introduction} \label{sec:introduction}

Diffuse clouds, being at the crossroad between the low density, mainly neutral and atomic gas phase, and the dense and fully molecular cloud phase where stars are born, not only trace this crucial transition, but play an active role in the lifecycle of gas in the interstellar medium (ISM). They provide the raw materials from which dense molecular clouds form. Being continuously replenished by feedback from late stages of star-formation and stellar evolution, diffuse clouds form a reservoir of heavy elements that is subsequently responsible for initiating the growth of chemical complexity in the ISM. Therefore, the study of diffuse clouds is crucial in advancing our interpretation of a variety of physical and chemical processes of broad applicability in astrophysics. These processes include those governing the transition from H- to H$_{2}$-dominated gas \citep{Federman1979, Sternberg2014, Bialy2016}, an essential ingredient in any complete theory of star formation and galaxy evolution. 

Being poorly shielded against photodissociating interstellar radiation fields, the low density ($n_{\rm H} \sim10^{2}$--$10^{3}$~cm$^{-3}$) gas comprising diffuse clouds was originally predicted to be devoid of molecules \citep{Eddington1926}. However, observations at ultraviolet, visible, and radio wavelengths have revealed an astounding inventory of molecules \citep{Goss1968, Whiteoak1970, Lucas1996, Snow2006, McGuire2021}. The surprisingly rich chemistry present in diffuse clouds can be attributed to exothermic reactions driven by cosmic-ray ionization and 
large fluctuations in the thermal pressure of the media. In more recent years, the molecular richness of diffuse clouds has been further explored 
by means of absorption spectroscopy at far-infrared (FIR) and sub-millimeter wavelengths against the background continuum emission of bright star–forming regions in both the Milky Way and external galaxies \citep[see][for a complete overview]{Gerin2016}. These observations -- largely performed with the \textit{Herschel} Space Observatory \citep[HSO,][]{Pilbratt2010}, the Atacama Pathfinder Experiment \citep[APEX,][]{Guesten2006}, and the Stratospheric Observatory for Infrared Astronomy \citep[SOFIA,][]{Young2012early} -- have not only extended absorption-line studies over Galactic scales, but have also renewed interest in the study of hydrides, defined here as molecules containing a single heavy element atom with one or more hydrogen atoms. 

Historically, hydrides have the distinction of being the first gas-phase molecules to be observed in the ISM, through the successful detection and identification of CH and CH$^+$ alongside CN at optical wavelengths near 4300~\AA\, \citep{dunham1937interstellar, swings1937, McKellar1940}. Thereafter, the ground-state transitions of OH near 18~cm became the first molecular signature to be detected in the ISM at radio wavelengths \citep{Weinreb1963}. Moreover, being amongst the first species to form in gas with both atomic and molecular hydrogen through relatively simple chemical pathways, hydrides play an important role in driving the chemistry of the ISM. 
Furthermore, because the FIR rotational transitions of most hydride species possess high rates of spontaneous radiative decay  and in turn high critical densities, hydride molecules are typically in the ground rotational state under diffuse cloud conditions ($n_{\rm H} \sim 10-500~$cm$^{-3}$); their measured absorption line strengths therefore yield robust and accurate estimates of the molecular column densities.

Observations performed over the past decade have allowed the ground-state rotational transitions of several interstellar hydrides to be observed for the first time at high spectral resolution 
and provided a wealth of new information and unique clues to our understanding of the physics of the diffuse ISM. These include sensitive tracers of atomic gas and the cosmic-ray ionization rate like ArH$^+$ \citep{Schilke2014, Jacob2020Arhp, Bialy2019}, OH$^+$, H$_{2}$O$^+$ and H$_{3}$O$^+$; \citep{ Gerin2010, Gupta2010, Neufeld2010a, Ossenkopf2010, Wyrowski2010, Indriolo2012w51, Indriolo2015}, molecular CO-dark H$_{2}$ gas tracers such as HF \citep{Neufeld2010, Sonnentrucker2015}, CH \citep{Gerin2010CH, Wiesemeyer2018, Jacob2019}, and OH \citep{Wiesemeyer2016}; and tracers of the dissipation of turbulence {or the turbulent transport at the interface of different phases} such as CH$^+$ and SH$^+$ \citep{Falgarone2010, Menten2011, Godard2014}. While the careful analysis of hydride abundances obtained from these observations have demonstrated the potential of absorption line studies of interstellar hydrides to probe the physical and chemical conditions in diffuse clouds, previous observations were limited to a fairly small set of sight lines almost entirely within the inner Galaxy.

 With the end of the \textit{Herschel} mission, SOFIA is the only observatory that can provide access that is almost unhindered by atmospheric absorption in the terahertz frequency range,
where the rotational transitions of most interstellar hydrides lie. Moreover, the GREAT\footnote{The German REceiver for Astronomy at Terahertz frequencies (GREAT) is a development by the MPI f\"{u}r Radioastronomie and the KOSMA/Universit\"{a}t zu K\"{o}ln, in cooperation with the DLR Institut f\"{u}r Optische Sensorsysteme.} instrument on SOFIA enables high resolution spectroscopy for the first time at frequencies that were inaccessible to the Heterodyne Instrument for the Far Infrared \citep[HIFI,][]{deGraauw2010} on \textit{Herschel}, permitting absorption line studies of the OH, SH, CH, and O transitions targeted in the HyGAL program.

Taking advantage of this unique capability, the HyGAL legacy survey aims to characterize the diffuse Galactic ISM, significantly expanding the set of observations available to probe the diffuse ISM throughout the Galaxy with observations of six key hydrides (ArH$^+$, p-H$_{2}$O$^+$, OH$^+$, SH, OH and CH) along with two atomic constituents of the diffuse ISM, (C$^+$ and O). The acquisition and analysis of these data are also very timely, as theoretical simulations of the turbulent multiphase ISM are starting to include chemistry, thereby allowing observations to be compared to specific model predictions for the first time. This paper presents an overview of the HyGAL program and the analysis of the first data collected from SOFIA Cycle 8 observations toward three of the targeted sources, W3(OH), W3~IRS5 and NGC~7538~IRS1. Section~\ref{sec:goals} details the main scientific goals of the HyGAL program, with the observations, use of archival and complementary data, while the data reduction strategies are described in Sect.~\ref{sec:observations}. We present and discuss the results in Sects.~\ref{sec:first_results} and \ref{sec:discussion},  and summarize our findings in Sect.~\ref{sec:summary}.

\section{HyGAL: Key science goals } \label{sec:goals}
The processes by which molecular clouds -- the sites of star formation, traced by CO -- form out of the atomic medium and the timescales on which molecular cloud formation occurs are crucial missing links in theories of star formation \citep{MacLow2004, McKee2007, Kennicutt2012, Padoan2014, Krumholz2019, Rosen2020}. Therefore, investigations of the diffuse atomic and molecular media are critical for our understanding of the formation and evolution of molecular clouds. While this topic is highly complex, the observations planned under the HyGAL legacy project will address several related questions about the ISM and star formation, including: 
\begin{itemize}[noitemsep]
    \item What is the distribution of the molecular fraction in various phases (diffuse atomic gas, diffuse molecular gas, and dense molecular gas) of the cold and warm neutral media of the ISM?
    \item How does the ionization rate due to low-energy cosmic-rays, an important source of heating in the ISM and responsible for regulating star-formation, vary within the Galaxy? 
    \item What is the nature of interstellar turbulence (e.g.\ the typical shear or shock velocities) and what mechanisms lead to its dissipation?
\end{itemize}
The following subsections detail the potential use of interstellar hydrides in addressing these key questions.

\subsection{Probing the transition from diffuse atomic to molecular gas }\label{subsec:diffuse_atomic_molecular}


 The molecular fraction, $f({\rm H}_{2})$, is a fundamental quantity that may be used to describe the transition between atomic and molecular gas within the cold neutral medium. It may be expressed either as a ratio of volume densities, $f^{n}({\rm H}_{2})$, or column densities, $f^{N}({\rm H}_{2})$:
\begin{align}
f^{n}({\rm H}_{2}) &= \frac{2n(\text{H}_{2})}{n(\text{H}) + 2n(\text{H}_{2})} \\
f^{N}({\rm H}_{2}) &= \frac{2N(\text{H}_{2})}{N(\text{H}) + 2N(\text{H}_{2})} \,,
\label{eqn:molfrac}
\end{align}
where $n(X)$ and $N(X)$ refer to the volume density (in units of cm$^{-3}$) and column density (in units of cm$^{-2}$) for $X = {\rm H, H}_{2}$, respectively.

Despite being the most abundant molecular species in the ISM, molecular hydrogen has been very difficult to observe directly, except through its ground electronic transitions near 1100~\AA\ at far-ultraviolet wavelengths in absorption \citep{Savage1977, Rachford2002, Rachford2009, Pan2005, Burgh2007, Shull2021} -- observations of which are restricted to bright nearby hot stars and toward bright QSOs -- and through its emission in infrared (IR) rovibrational transitions \citep{Burton1992, Habart2011, Santangelo2014}. Rovibrational emission from H$_2$ is typically excited in shock-heated or UV-irradiated regions, so UV absorption lines observed toward nearby hot stars provide the only direct probe of the cold H$_2$ in the ISM. Because it is homonuclear and symmetric, H$_{2}$ does not possess a permanent dipole moment and radiates only through weak quadrupole transitions of its excited rotational and rovibrational states at wavelengths below $\sim28~\mu$m ($h\nu/k_{\rm B}\sim500~$K). Furthermore, with high-lying and widely spaced energy levels the rotational excitation of H$_2$ becomes important only for gas temperatures $T\geq80$~K, making infrared H$_2$ emissions from the cold ISM unobservable in practice. The challenges involved in observing molecular hydrogen are overcome by using chemically associated species and dust as surrogates for H$_2$. While the integrated intensity of low-lying rotational transitions of CO scaled by an empirical conversion factor $X$(CO) has been the most commonly used proxy for measuring the total molecular reservoir, there is mounting evidence for the presence of excess molecular material not traced by CO, dubbed as CO-dark H$_{2}$ gas \citep{Grenier2005}, which may constitute up to 50~\% of the total mass of the molecular gas along any given sight line \citep{Pineda2013}. This excess molecular gas points to the inability of CO to act as a proxy for H$_2$, in regions of low dust column densities (or visual extinction) where CO readily undergoes photodissociation. In recent years, through observations of their radio and FIR rotational transitions, hydrides such as CH \citep{Gerin2010, Wiesemeyer2018, Jacob2019}, OH \citep{Allen2015, Engelke2018, Rugel2018,Busch2019, Busch2021} and HF \citep{Neufeld2010, Sonnentrucker2015} have been established as important tracers of CO-dark H$_{2}$ gas in addition to the [C\,{\small II}] 158~$\mu$m line \citep{Langer2014} and the $J=1-0$ rotational transition of HCO$^+$ \citep{Lucas1996, Gerin2010CH}. We note that the CH--H$_{2}$ relationship was first established by \citet{Federman1982} and later by \citet{Sheffer2008} and then \citet{Weselak2019} using optical observations of the $\text{A}^2\Delta$ - $\text{X}^2\Pi$ system of CH at $4300\,$\AA\ and the (2$-$0), (3$-$0), and (4$-$0) bands of the Lyman B-X transitions of H$_{2}$ toward stars located in the local diffuse ISM ($10^{19} < N(\text{H}_{2})<10^{21}\,$cm$^{-2}$). The
validity of this relationship between CH and H$_{2}$ over Galactic scales is poorly known, having been determined only using a sample of nearby stars that are bright at visible wavelengths, and might be dependent on variations in the elemental abundance of carbon across the Galaxy. Nevertheless, a linear CH -- H$_{2}$ relationship with a constant slope has been widely adopted in previous studies, an approach we follow here.

\begin{figure}[!t]
    \centering
    \includegraphics[width=0.5\textwidth]{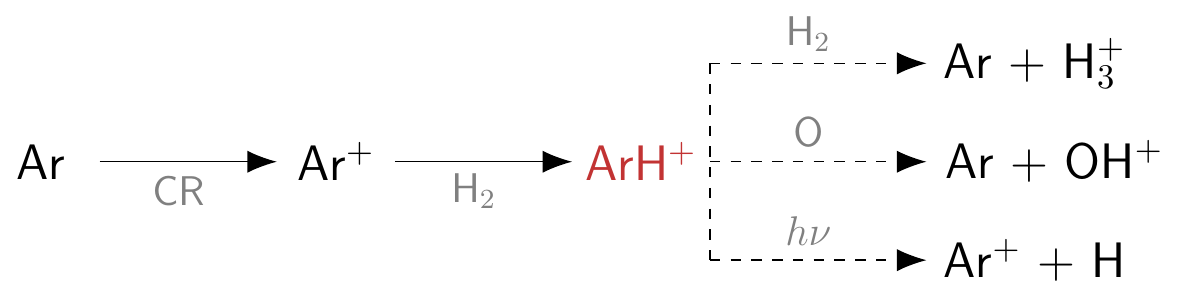}
    \caption{Schematic representation of the argon chemistry in diffuse clouds. Solid arrows display chemical pathways leading to the formation of ArH$^+$ initiated by cosmic-rays (CR) while the dashed arrows indicate competing reactions that reduce the abundance of this cation. 
 }
    \label{fig:argon_chemistry}
\end{figure}

\begin{figure*}[!t]
    \centering
    \includegraphics[width=0.7\textwidth]{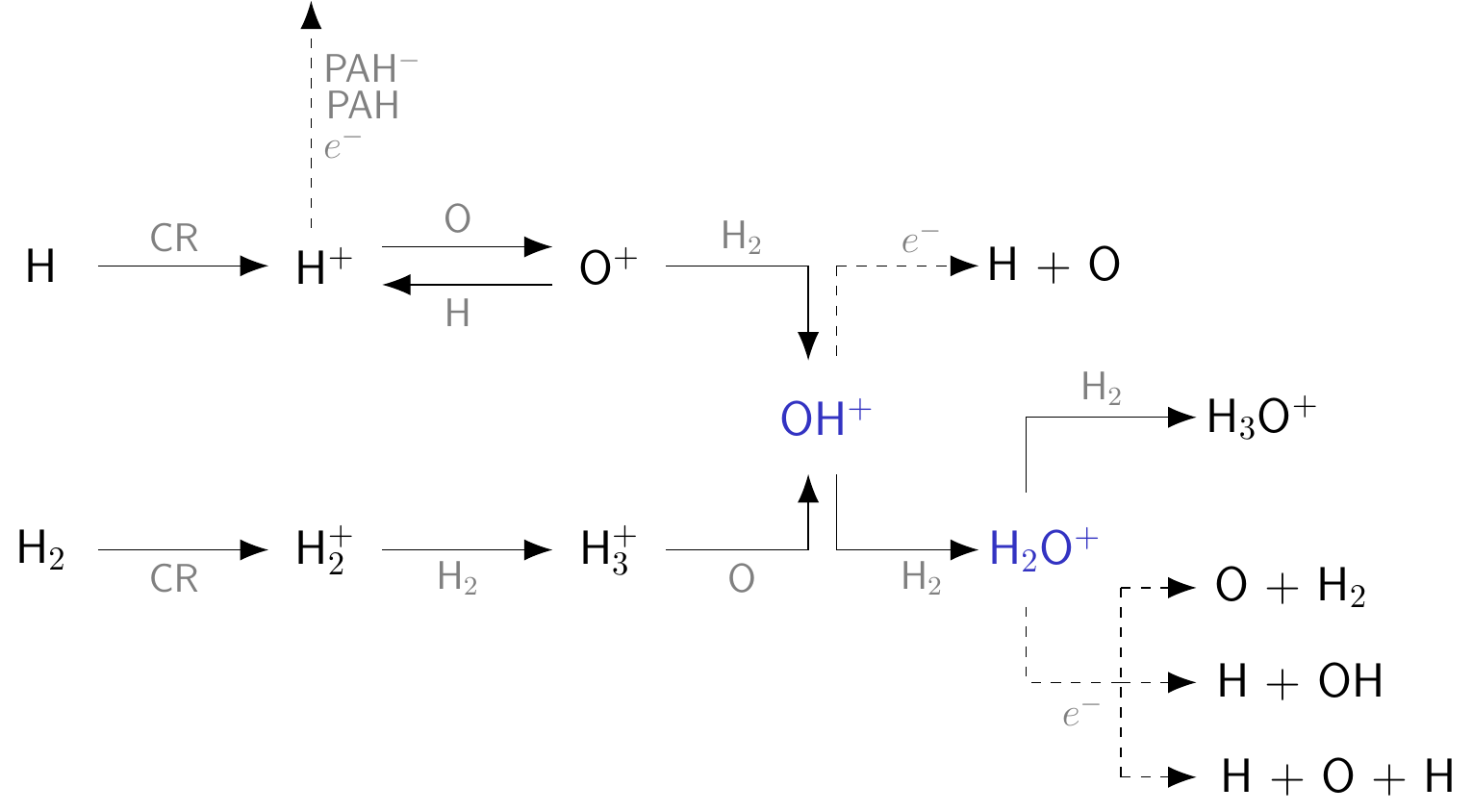}
    \caption{Schematic representation of the oxygen chemistry in diffuse clouds. Solid arrows display chemical pathways leading to the formation of oxygen hydride cations-- OH$^+$ and H$_2$O$^+$ while the dashed arrows indicate competing reactions that reduce the abundance of these cations.
 }
    \label{fig:oxygen_chemistry}
\end{figure*}

The neutral hydrides that are observed in the HyGAL project -- CH, OH, and SH -- are known to show peak abundances in gas with a relatively large fraction of hydrogen in molecular form ($f^{n}({\rm H}_{2}) > 0.1$), tracing diffuse molecular and translucent clouds \citep[as per the classification of the neutral medium by][]{Snow2006}. In contrast, the hydride ions targeted in this study -- ArH$^+$, p-H$_{2}$O$^+$, and OH$^+$ -- are rapidly destroyed by reactions with H$_2$ and are therefore most abundant in material with molecular fractions of a few percent (see Figures~\ref{fig:argon_chemistry} and \ref{fig:oxygen_chemistry} and Figure 1 of \citet{Bialy2019}). 
The $n$(OH$^+$)/$n$(H$_2$O$^+$) ratio is a particularly valuable probe of the H$_2$ fraction \citep{Neufeld2010a} because the reaction between H$_2$ and OH$^+$, which forms H$_2$O$^+$, competes with the dissociative recombination of OH$^+$ (Figure~\ref{fig:oxygen_chemistry}); thus, the $n$(OH$^+$)/$n$(H$_2$O$^+$) ratio is larger in regions of small $f^{n}$(H$_2$) and smaller in regions of large $f^{n}$(H$_2$). 

An analysis of OH$^+$ and H$_2$O$^+$ absorption features observed using \textit{Herschel} along sight lines toward bright Galactic continuum sources \citep{Indriolo2015} showed these molecules to trace H$_{2}$ fractions of a few percent ($f^{N}$(H$_2$) =10$^{-2}$--10$^{-1}$). However, having been determined using ratios of column densities rather than volume density (i.e. $N$(OH$^+$)/$N$(H$_2$O$^+$) rather than $n$(OH$^+$)/$n$(H$_2$O$^+$)), the derived molecular fractions only represent averages across the entire region where the OH$^+$ and H$_2$O$^+$ absorption features
are produced. Therefore, for any given sight line, these $f^{N}$(H$_2$) values may differ substantially from those derived using measurements of $n$(H$_2$) and $n$(H). The argonium ion ArH$^+$, on the other hand, traces material of even lower molecular fraction ($f^{N}$(H$_2)=10^{-4}$--10$^{-2}$) and high cosmic-ray ionization rates, which paradoxically makes it a better tracer of almost purely atomic hydrogen gas than H\,{\small I} itself, as discussed in \citet{Schilke2014} and \citet{Neufeld2016}. 
Taken together, the observations of both the ionized and neutral hydride species carried out in the HyGAL project trace different cloud layers across any given sight line and reveal the transition from diffuse atomic ($n_{\rm H} = 10-100~{\rm cm}^{-3}; f^{n}({\rm H}_{2})<0.1$) to diffuse molecular ($n_{\rm H} = 100-500~{\rm cm}^{-3}; f^{n}({\rm H}_{2})>0.1$) and translucent clouds \citep[$n_{\rm H} = 500-5000~{\rm cm}^{-3}; f^{n}({\rm H}_{2})\sim0.8$,][]{Snow2006}.

\subsection{Probing the cosmic-ray ionization rate in the diffuse ISM}\label{subsec:crir}
In addition to providing valuable probes of the molecular fraction in the Galactic ISM, the observed abundances of the hydride ions OH$^+$, H$_2$O$^+$, and ArH$^+$ allow us to estimate the cosmic-ray ionization rates. Since the first ionization potentials of oxygen (13.62~eV) and argon (15.76~eV) are higher than that of hydrogen (13.59~eV), both atoms remain primarily in their neutral states in the cold neutral medium as they are well-shielded from extreme-ultraviolet radiation capable of ionizing them. Therefore, the formation of oxygen-, and argon-bearing hydride ions such as OH$^+$, H$_2$O$^+$, and ArH$^+$ is driven by cosmic-ray ionization \citep{Gerin2010, Neufeld2010a, Neufeld2016}. The formation of ArH$^+$ ensues via the cosmic-ray ionization of Ar to produce Ar$^+$, which quickly reacts with H$_2$ to form ArH$^+$ (see Figure~\ref{fig:argon_chemistry}) while, oxygen chemistry in diffuse clouds, initiated by the cosmic-ray ionization of H or H$_2$, follows a sequence of ion-neutral reactions resulting in the formation of OH$^+$ and H$_{2}$O$^+$ as shown in Figure~\ref{fig:oxygen_chemistry}. 


Carrying out a detailed analysis of data available from \textit{Herschel} on OH$^+$, H$_2$O$^+$, and ArH$^+$, \citet{Neufeld2017} obtained an estimate of the cosmic-ray ionization rate within diffuse atomic clouds with an average value $\zeta_{\rm p}(\text{H})=(2.2\pm0.3)\times10^{-16}$~s$^{-1}$. A similar value is obtained by \citet{Jacob2020Arhp} using ArH$^+$ data collected using the APEX 12~m telescope. While this value is in excellent agreement with entirely independent estimates of $\zeta_{\rm p}(\text{H})$ obtained for diffuse molecular clouds (with $f( \text{H}_{2})>0.1$) using observations of H$_3^+$ \citep{McCall2003, Indriolo2007, Indriolo2012}, it is an order of magnitude larger than the typical cosmic-ray ionization rates inferred for dense molecular clouds \citep{Caselli1998, Bialy2021} and expectations based on direct measurements of cosmic-rays obtained with the Voyager~I spacecraft \citep{Cummings2016}. This suggests that cosmic-rays are excluded from dense molecular clouds \citep[e.g.][]{Padovani2009, Padovani2020} and that the cosmic-ray flux in the vicinity of the solar system may be atypical. Furthermore, results obtained from existing H$_{3}^+$ data by \citet{Neufeld2017} suggest at a marginal level of significance that the cosmic-ray ionization rates in diffuse molecular clouds decrease with cloud extinction for $A_{\rm V} \geq 0.5$. 

Observations of the key cosmic-ray ionization rate tracers, OH$^+$, H$_{2}$O$^+$ and ArH$^+$ to be performed in the  HyGAL project  --  in combination with ancillary observations (discussed further in Sect.~\ref{subsec:complementary_data}) of tracers of higher total gas column and molecular fraction such as HCO$^+$ and CO -- will not only more than double the sample of Galactic sight lines toward which the cosmic-ray ionization rate has been determined, but will also provide better statistics to understand the nature of the cosmic-ray ionization rates in different cloud environments. 

\subsection{Probing the chemistry of warm phases of the ISM}
While the chemistry of OH$^+$, H$_{2}$O$^+$, and ArH$^+$ discussed in Sects.~\ref{subsec:diffuse_atomic_molecular}, and \ref{subsec:crir} proceed via exothermic reactions that are rapid at temperatures typical of the diffuse ISM \citep[$\leq$100~K,][]{Snow2006}, that of other hydrides and in particular those containing sulfur (like SH, SH$^+$, H$_{2}$S) and CH$^+$, are distinctive in that their formation only takes place by means of endothermic reactions.
The latter may be driven by elevated gas temperatures or 
ion-neutral drift in shocks associated with turbulent dissipation regions; or via turbulent transport at the interface between the cold and warm neutral media \citep{Meyers2015, Valdivia2017, Moseley2021}. 

Observationally, SH was first detected using the GREAT instrument by \citet{Neufeld2012} in absorption toward the star-forming region W49N and subsequently detected toward a handful of other star-forming regions \citep{Neufeld2015S}. The measured abundances of SH in the diffuse ISM are found to be many orders of magnitude larger than those predicted for quiescent, cold clouds by standard models of photodissociation regions (PDRs), suggesting that some fraction of the volume is occupied by a warmer phase heated by shocks or other processes that lead to the dissipation of interstellar turbulence \citep{Godard2009, Godard2014, Neufeld2015S}. Similarly, \citet{Walch2015} have found using the SILCC\footnote{SILCC- SImulating the LifeCycle of molecular Clouds} simulations that the shock heating of a large fraction of gas which neighbors molecular clouds lies above the thermal equilibrium curve thereby pushing this material into the thermally unstable regime. While such modeling efforts facilitate the interpretation of the chemistry of sulfur-bearing hydrides, the abundance estimates are based on a limited number of detections in diffuse clouds. 

Observations of SH performed in the HyGAL program will provide high-quality spectra toward individual targets probing different sight lines. In combination with ancillary data on other small molecules containing sulfur, such as CS and H$_{2}$S, they will provide key information about the overall sulfur chemistry in the diffuse and translucent clouds. Interpreted in the context of chemical models of shocks and turbulent dissipation regions, the observed abundances of sulfur-bearing molecules further promises to provide important constraints on the nature of interstellar turbulence.

\label{subsec:sh_chem}

\section{Observations} \label{sec:observations}
\subsection{Source selection} \label{subsec:sourceselection}

The goals of the HyGAL project discussed in Sect.~\ref{sec:goals} are met by performing absorption-line spectroscopy of foreground interstellar gas along the sight lines to bright background continuum sources. The most useful sight lines intersect as many spiral arms as possible, including those in the outer Galaxy (i.e., sources located outside the solar circle at Galactocentric distances, $R_{\rm GAL} >8.15$~kpc), to probe the structure of the ISM. Our main selection criterion, however, is driven by sensitivity. Since the signal-to-noise ratio obtained for an absorption feature is proportional to the continuum flux for a given column density, the most favorable sources are those possessing the largest continuum fluxes within the far-IR wavelength range (119--494~$\mu$m) encompassing the targeted line transitions (see, Table~\ref{tab:spectroscopic_properties}). Suitable background continuum sources were identified using the source catalog of the Hi-GAL \textit{Herschel} key-guaranteed time project \citep{Molinari2010}, which provides continuum fluxes at 70, 160, 250, 350 and 500~$\mu$m. The set of sources targeted in this survey consists of 25 submillimeter continuum sources in the Galactic plane with 160~$\mu$m (1.9~THz) continuum fluxes in excess of 2000~Jy for sources in the inner Galaxy and 1000~Jy for those in the outer Galaxy, as estimated from the Hi-GAL source catalog \citep{Elia2021}. The positions of the background sources were checked against known millimeter-wave sources, H{\small II} regions and masers using the SIMBAD\footnote{See, \href{http://simbad.u-strasbg.fr/simbad/}{http://simbad.u-strasbg.fr/simbad/}.} astronomical database \citep{Wenger2000}. In addition, the HyGAL source list also includes a few sources for which partial spectroscopic data is available from \textit{Herschel}/HIFI, most often collected as a part of the PRISMAS\footnote{PRobing InterStellar Molecules with Absorption line Studies \citep[PRISMAS,][]{Gerin2010}.} guaranteed time key program, but where the complete set of species targeted in HyGAL has not been observed. Therefore, the completion of this legacy program will roughly double the number of sight lines along which the hydrides targeted in this survey have been observed; equally importantly, it will extend observations toward regions in the outer Galaxy that were previously unexplored. The HyGAL sources are situated over a wide range of Galactic longitude, lying at estimated distances from the Sun ranging from 1.3 to 13~kpc. 

\begin{figure}
    \includegraphics[width=0.49\textwidth]{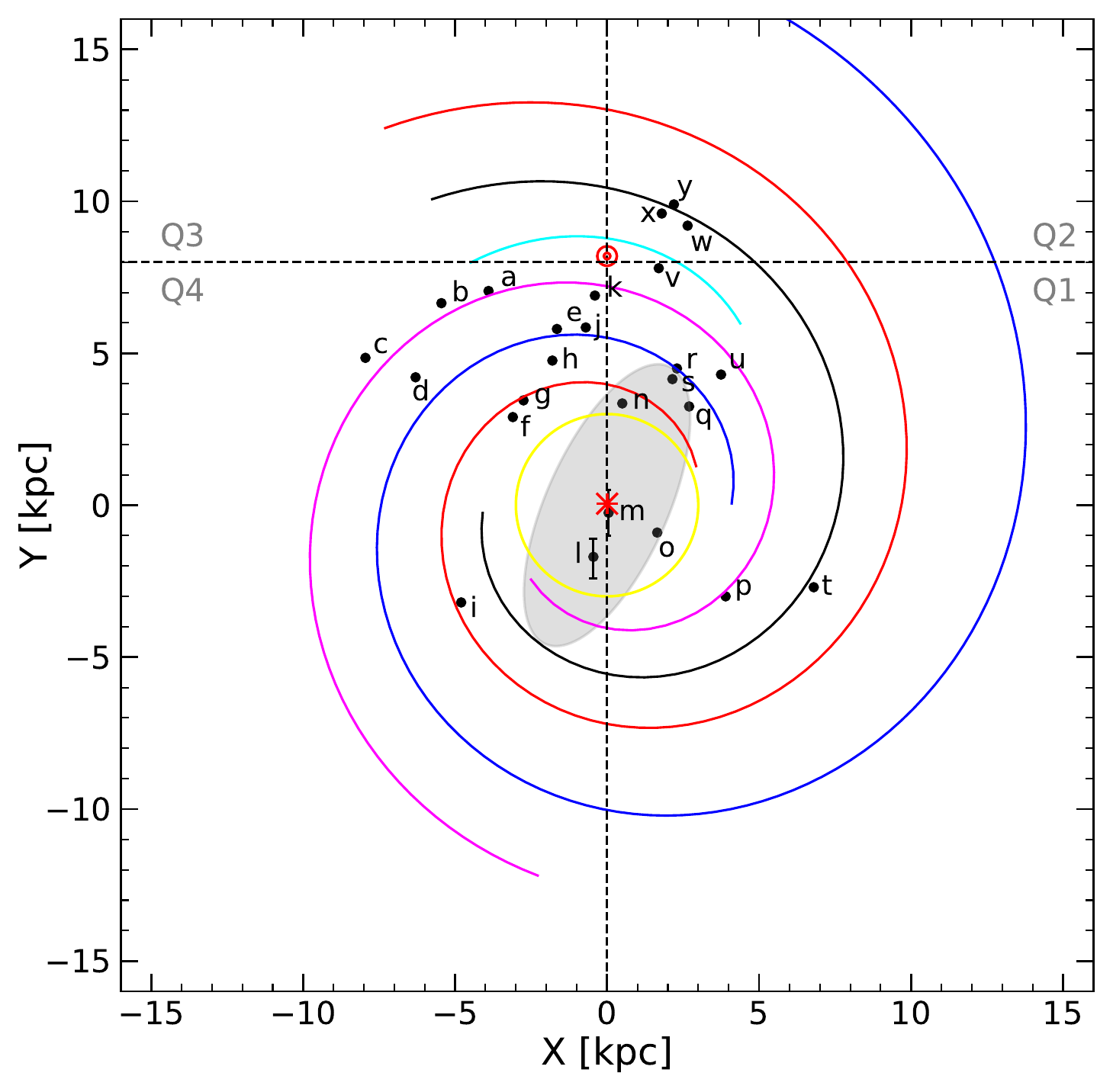}
    \caption{Distribution of the sight lines observed under the HyGAL program (black circles), as viewed from the north Galactic pole. The underlying trace of the Milky Way spiral arm pattern follows the log-periodic model presented in \citet{Reid2019}; the Galactic center (red asterisk) is at (0, 0)~kpc and the Sun (red Sun symbol) is at (0, 8.15)~kpc; 3~kpc ring, yellow; Norma–Outer arm, red; Scutum–Centaurus–OSC arm, blue; Sagittarius–Carina arm, purple; Local arm, cyan; Perseus arm, black; the `long' bar is indicated by a shaded ellipse after \citet{Wegg2015}. The source names are designated using alphabetic letters (as given in Table~\ref{tab:hygal_complete_source_list}) for clarity, five of which are located in the outer Galaxy. The sources labelled as l and m in this plot are depicted using bars due to uncertainties in their kinematic distances. In this paper we discuss sources labelled w, x and y, respectively.}
    \label{fig:source_overview}
\end{figure}

Table~\ref{tab:hygal_complete_source_list} provides an overview of the properties of the HyGAL source sample, and Figure~\ref{fig:source_overview} shows their distribution in the Galactic disk. Furthermore, a brief description of each source and its line-of-sight properties is given in Appendix~\ref{sec:los_source_properites}.


\begin{table*}
    
     \caption{Spectroscopic properties of the studied species and transitions.}
    \begin{center}
    \begin{tabular}{lccccrc rl }
    \hline \hline
    Species & \multicolumn{2}{c}{Transition} & Frequency & $A_{\text{u,l}}$ & \multicolumn{1}{c}{$E_{\text{u}}$} & GREAT  & HPBW & $\eta_{\rm MB}$\tablenotemark{a}\\
    &  $J^{\prime} - J^{\prime\prime}$ & $F^{\prime} - F^{\prime\prime}$ & [GHz] & [s$^{-1}$] & \multicolumn{1}{c}{[K]} & \multicolumn{1}{c}{channel} & \multicolumn{1}{c}{[$^{\prime\prime}$]} & \\
    \hline 
    
    \arhp  & $1 - 0$  & --- & 617.5252(2) & 0.0045 & 29.63 & 4G1 & 44.7 & 0.61\\
   
    p-\htop & $3/2 - 1/2$& --- & 604.6841(8)& 0.0013 & 29.20 &  4G1 & 44.7 & 0.61\\ 
    ~~$N_{K_{\rm a}K_{\rm c}} = 1_{1,0} - 1_{0,1}$     & $3/2 - 3/2$&  --- & 607.2258(2) & 0.0062 & 29.20 &\\  
     \ohp & $2-1$ & $5/2 - 3/2$ & 971.8038(15)\tablenotemark{b} & 0.0182 & 46.64 &4G2 & 28.8 & 0.50 \\ 
    ($N$=1$-$0)& & $3/2 - 1/2$ & 971.8053(15) & 0.0152 & \\
         & & $3/2 - 3/2$ & 971.91920(10) & 0.0030 & \\ 
    SH & $5/2 - 3/2$ & $2^{+} - 2^{-}$ & 1382.9040(1) & 0.0005 & 66.40 & 4G3 & 19.3& 0.60\\
     ~~$^2\Pi_{3/2}, N = 2$ &  & $3^{+} - 2^{-}$  & 1382.9086(1) & 0.0047 & & \\
     & & $2^{+} - 1^{-}$ & 1382.9152(1) & 0.0042 & & \\
     & & $2^{-} - 2^{+}$ & 1383.2350(1) & 0.0005 &  &  & \\
      &  & $3^{-} - 2^{+}$  & 1383.2397(1) & 0.0047 & & \\
     & & $2^{-} - 1^{+}$ & 1383.2462(1)\tablenotemark{b} & 0.0042 & & \\
    
    OH & $5/2-3/2$ & $2^{-} - 2^{+}$ & 2514.2987(9) & 0.0137 & 120.75 & 4G4 & 11.2 & 0.52\\
    ~~$^2\Pi_{3/2}, N = 2-1$ & & $3^{-} - 2^{+}$ & 2514.3167(9)\tablenotemark{b} & 0.1368\\
    & & $2^{-} - 1^{+}$ & 2514.3532(9) & 0.1231 \\
     CH &  $3/2-1/2$ & $1^{-}-1^{+}$ & 2006.74886(6) &  0.0111& 96.31 & LFA & 13.5 & 0.66\\
      ~~$^2\Pi_{3/2}, N = 2-1$ & & $1^{-}-0^{+}$ & 2006.76258(6) & 0.0223 & \\
            & & $2^{-}-1^{+}$ & 2006.79906(6)\tablenotemark{b} & 0.0335 &  \\
     \cii  $\quad {}^{2}P_{3/2}-{}^{2}P_{1/2}$& $2-1$ & --- & 1900.5369(13) & $2.32\times10^{-6}$ & 91.21& LFA & 14.1 & 0.66\\
     \oi  $\quad {}^{3}P_{1}-{}^{3}P_{2}$& $2-1$ & --- & 4744.7775(1) & $8.91\times10^{-5}$& 227.76 & HFA & 6.3 & 0.65\\
            
            \hline    
     \end{tabular}
     \end{center}
     \tablecomments{ The spectroscopic data are taken from the Cologne Database for Molecular Spectroscopy \citep[CDMS,][]{muller2005cologne}. The H$_2$O$^+$ frequencies were refined considering astronomical observations \citep[see Appendix A of][]{Mueller2016} for which the upper level energies are given with respect to the ground state of p-H$_2$O$^+$ ($N_{K_{\rm a}K_{\rm c}} = 1_{0,1}$) and the CH frequencies are taken from \citet{Truppe2014}. For the rest frequencies, the numbers in parentheses give the uncertainty in the last listed digit.} \tablenotetext{a}{The main-beam efficiencies for each channel (and pixel where applicable) are determined for each flight series, listed here are the typical values.} \tablenotetext{b}{Indicates the hyperfine structure transition that was used to set the velocity scale in the analysis.}
 \label{tab:spectroscopic_properties}
\end{table*}

\begin{table*}
{
   \begin{center}
        \caption{HyGAL source parameters. }
        \label{tab:hygal_complete_source_list}
    \begin{tabular}{ll cc rr r c r l}
    \hline\hline
        \# &Source & Right Ascension & Declination & Gal. Long. & Gal. Lat.  & $\upsilon_{\rm LSR}$ & \multicolumn{1}{c}{$d$~[Ref]} & $R_{\rm GAL}$  \\
        &Designation & [hh:mm:ss] & [dd:mm:ss] & \multicolumn{1}{c}{[deg]} & \multicolumn{1}{c}{[deg]} & [km~s$^{-1}$]& [kpc] & [kpc] \\
         \hline 

a & HGAL284.015$-$00.86 &	10:20:16.1	& $-$58:03:55.0 & 284.016  & $-$0.857  & 9.0 & 5.7~[1]& 9.0\\ 
b & HGAL285.26$-$00.05  &	10:31:29.5	& $-$58:02:19.5 & 285.263  & $-$0.051  & 3.4 & 4.3~[2] & 8.2\\ 
c & G291.579$-$00.431   &	11:15:05.7	& $-$61:09:40.8 & 291.579  & $-$0.431  & 13.6 & 8.0~[3] & 9.3\\ 
d & IRAS~12326-6245     &	12:35:35.9	& $-$63:02:29.0 & 301.138  & $-$0.225  & $-$39.3 & 4.6~[4] & 7.2\\ 
e &G327.3$-$00.60 	    &	15:53:05.0	& $-$54:35:24.0 & 327.304  & $-$0.551  & $-$46.9 & 3.1~[5] & 6.2\\ 
f & G328.307+0.423 	    &	15:54:07.2	& $-$53:11:40.0 & 328.309  & +0.429    & $-$93.6  & 5.8~[6] & 4.6\\ 
g & IRAS~16060$-$5146   &	16:09:52.4	& $-$51:54:58.5 & 330.953  & $-$0.182  & $-$91.2 & 5.3~[7] & 4.5\\ 
h & IRAS~16164$-$5046   &	16:20:11.9	& $-$50:53:17.0 & 332.827  & $-$0.551  & $-$57.3 & 3.6~[8] & 5.4\\ 
i & IRAS~16352$-$4721   &	16:38:50.6	& $-$47:28:04.0 & 337.404  & $-$0.403 & $-$41.4  & 12.3~[4] & 5.1\\ 
j & IRAS~16547$-$4247   &	16:58:17.2	& $-$42:52:08.9 & 343.126  & $-$0.063  & $-$30.6 & 2.7~[8] & 5.8\\ 
k & NGC~6334~I  	    &	17:20:53.4	& $-$35:47:01.5 & 351.417  & +0.645    & $-$7.4  & 1.3~[9] & 7.0\\ 
l & G357.558$-$00.321   &	17:40:57.2	& $-$31:10:59.3 & 357.557  & $-$0.321  & 5.3  & 9.0--11.8~[10] & 1.0--3.6\\
m & HGAL0.55$-$0.85     &	17:50:14.5	& $-$28:54:30.7 & 0.546    & $-$0.851  & 16.7 & 7.7--9.2~[11] & 0.4--1.0 \\
n & G09.62+0.19         &	18:06:14.9	& $-$20:31:37.0 & 9.620    & +0.194    & 4.3  & 5.2~[12] & 3.3 \\
o & G10.47+0.03         &	18:08:38.4	& $-$19:51:52.0 & 10.472   & +0.026    & 67.6    & 8.6~[13] & 1.6 & \\
p & G19.61$-$0.23       &	18:27:38.0	& $-$11:56:39.5 & 19.608   & $-$0.234  & 40.8    & 12.6~[14] & 4.7 \\
q & G29.96$-$0.02       &	18:46:03.7	& $-$02:39:21.2 & 29.954   & $-$0.016  & 97.2    & 6.7~[6] & 4.5 \\
r & G31.41+0.31         &	18:47:34.1	& $-$01:12:49.0 & 31.411   & +0.307    & 98.2    & 4.9~[15] & 5.0 \\
s & W43~MM1             &	18:47:47.0	& $-$01:54:28.0 & 30.817   & $-$0.057  & 97.8    & 5.5~[15] &5.0 \\
t & G32.80+0.19         &	18:50:30.6	& $-$00:02:00.0 & 32.796   & +0.191    & 14.6    & 13.0~[16] & 7.4 \\
u & G45.07+0.13         &	19:13:22.0	& +10:50:54.0   & 45.071   & +0.133    & 59.2    & 4.3~[17] & 6.2 \\
v & DR21                &   20:39:01.6  & +42:19:37.9   & 81.681   & 0.537     & $-$4.0 & 1.5~[18] & 7.4 \\
\textbf{w} & \textbf{NGC~7538~IRS1}       &   \textbf{23:13:45.3}  & \textbf{+61:28:11.7}   & \textbf{111.542}  & \textbf{0.777}     & \textbf{$\boldsymbol{-}$59.0} & \textbf{2.6}~\textbf{[19]} & \textbf{9.8} \\
\textbf{x} & \textbf{W3~IRS5 }            &   \textbf{02:25:40.5}  & \textbf{+62:05:51.0}   & \textbf{133.715}  & \textbf{1.215}     & \textbf{$\boldsymbol{-}$39.0} & \textbf{2.3}~\textbf{[20]} & \textbf{9.9}\\
\textbf{y} & \textbf{W3(OH)}              &   \textbf{02:27:04.1}  & \textbf{+61:52:22.1}   & \textbf{133.948}  & \textbf{1.064}     & \textbf{$\boldsymbol{-}$48.0} & \textbf{2.0}~\textbf{[20, 21]} & \textbf{9.6} & \\
         \hline 
    \end{tabular}
    \end{center}
    \tablecomments{
    Those entries highlighted in bold text refer to the sources discussed in this work the remaining sources have either only recently been observed or are yet to be observed.}
\tablerefs{ For the heliocentric distances:
[1]~\citet{Urquhart2014a}; [2]~\citet{Caswell1995}; [3]~\citet{Lee2012}; [4]~\citet{Green2011}, [5]~\citet{Wienen2015}; [5]~\citet{Urquhart2018}; [7]~\citet{Moises2011}; [8]~\citet{Giannetti2014}; [9]~\citet{Immer2013}; [10]~\citet{Frail1996} (Lacking distance estimates, for G357.558$-$00.321 we adopt a distance equivalent to the nearby supernova remnant G357.7$-$0.1 (MSH 17$-$39); [11]~\citet{Walsh1998}; [12]~\citet{Sanna2009}; [13]~\citet{Sanna2014}; [14]~\citet{Urquhart2014}; [15]~\citet{Zhang2014} (For G31.41+0.31, following \citet{Winkel2017} we adopt a distance which is the average between two sources with similar LSR velocities, G31.28+0.06 and G31.58+0.07 for which accurate distances were measured by \citet{Zhang2014} and consistent with \citet{Reid2017}.); [16]~\citet{Kolpak2003}; [17]~\citet{Stead2010}; [18]~\citet{Rygl2012}; [19]~\citet{Moscadelli2009}; [20]~\citet{Navarete2019}; [21]~\citet{Hachisuka2006}. These distances were also cross-checked with those recently calculated by \citet{Mege2021}.}
}
\end{table*}

\subsection{SOFIA observations} \label{subsec:sofiadata}
The science goals discussed in Sect.~\ref{sec:goals} are achieved by exploiting the complete capabilities of the high resolution heterodyne instrument, GREAT on board SOFIA, which provides six frequency bands of which up to five can be used simultaneously. The observations were carried out over several flights as a part of the observatory's Cycle 8, and the ongoing Cycle 9 campaigns, under the open time project 08\_0038. The receiver configuration consists of the 4GREAT module \citep{Duran2020} whose different frequency channels are tuned to observe five key hydrides (\arhp, \htop, \ohp, SH, and OH), along with the High Frequency Array (HFA) of upGREAT \citep{Risacher2018}, tuned to the 63~$\mu$m fine-structure line of atomic oxygen. Additional configurations employ the Low Frequency Array (LFA) to observe CH or the 158~$\mu$m [\cii] line. The spectroscopic parameters of the different transitions studied and receiver configurations used are presented in Table~\ref{tab:spectroscopic_properties}.

The spectra are taken in the double-beam switch mode, chopping at a frequency of 2.5~Hz with a chop throw between 180$^{\prime\prime}$ and 300$^{\prime\prime}$ and a chop angle between $-10^{\circ}$ and 70$^{\circ}$ (East of North) to account for both atmospheric and instrumental fluctuations that may arise. The receiver is connected to a modified version of the MPIfR Fast Fourier Transform Spectrometer described by \citet{klein2012}. This backend provides a $4\,$GHz bandwidth per frequency band, per spatial position (equivalent to velocity widths of $\sim$1945, 1235, 868, 478, 615 and 252~km~s$^{-1}$ for the 4G1, 4G2, 4G3, 4G4, LFA and HFA receivers, respectively) over 16384 channels. This implies a native velocity resolution of 0.118, 0.075, 0.053, 0.029, 0.037 and 0.015~km~s$^{-1}$ for the 4G1, 4G2, 4G3, 4G4, LFA and HFA receivers, respectively. The high resolution thereby achieved makes it possible to separate the spiral arms from the envelopes of the hot cores.\\

The 
4G1 channel is tuned to observe the 607.225~GHz $N_{K_{a},K_{c}} = 1_{1,0}-1_{0,1}$ line of p-\htop\ in the lower sideband (LSB) and the 617.525~GHz $J=1-0$ transition of \arhp in the upper sideband (USB). Previous observations of the 607.225~GHz p-\htop\ line absorption spectra \citep{Schilke2010, Jacob2020Arhp} clearly show blended emission features at the systemic velocity of the star-forming regions studied: these arise from the H$^{13}$CO$^{+}$ ($J = 7-6$) and CH$_{3}$OH $J_k = 12_2-11_1~\text{E}$ lines at 607.1747 \citep{Lattanzi2007} and 607.2159~GHz \citep{Belov1995}, respectively. The line-of-sight also contains the D$_{2}$O ($J_{K_a,K_c} = {1_{1,1}-0_{0,0}}$) transition at 607.349~GHz \citep{matsushima2001frequency}, although this transition
has so far only been observed towards the solar type protostar IRAS16293-2422 \citep{Butner2007,Vastel2010} and is unlikely to result in significant contamination. Therefore, in order to decompose contributions from both sidebands and disentangle any additional contamination present in the bandpass, two additional setups are used with offsets in the LO frequency corresponding to Doppler velocities of $\pm$15~km~s$^{-1}$. A 15~km~s$^{-1}$ shift in the signal band corresponds to a 30~km~s$^{-1}$ relative shift in the image band. This observing strategy therefore results in a net image band offset of 60~km~s$^{-1}$ among the different setups, which is sufficient to separate features arising in the signal band from those in the image band. Furthermore, for the LFA observations only data collected from the central pixels of both polarizations are used because the continuum flux is much weaker or almost undetectable in the outer pixels which are separated from the central pixel by $\sim 31.8^{\prime\prime}$. 

The data is processed using the KOSMA kalibrate program \citep{guan2012great} and then converted to main-beam temperature scales, $T_{\rm MB}$, assuming a forward efficiency, $\eta_{\rm F}$, of 0.97 and main-beam efficiencies, $\eta_{\rm MB}$, listed in Table~\ref{tab:spectroscopic_properties}. The calibrated spectra are subsequently processed using the GILDAS-CLASS software\footnote{Software package developed by IRAM, see \url{https://www.iram.fr/IRAMFR/GILDAS/} for more information regarding GILDAS packages.} and low order ($\leq 2$) polynomial baselines are removed. The spectra are further re-sampled to a uniform channel width of 0.5~km~s$^{-1}$ (except for the OH$^+$ line toward NGC~7538~IRS1, for which the spectra are displayed with a velocity resolution of 1.5~km~s$^{-1}$ for clarity).

As of September 2021, 39.4~\% ($\approx 32~$hours) of the total planned observations for the HyGAL Legacy program have been performed. Table~\ref{tab:sofia_obs_update} summarizes the current status of the observations toward each source. The post-processed data products delivered to the SOFIA Science Center by the GREAT team will be publicly accessible via the NASA/IPAC InfraRed Science Archive\footnote{\href{https://irsa.ipac.caltech.edu/applications/sofia/}{https://irsa.ipac.caltech.edu/applications/sofia/}.} under Plan ID 08\_0038. 

\begin{table*}[]
    \begin{center}
    \caption{Status of HyGAL SOFIA survey as of September 2021.}
    \label{tab:sofia_obs_update}
    \begin{tabular}{l lllll lll}
    \hline\hline
    Source & \multicolumn{8}{c}{Species}\\
    & ArH$^+$ & p-H$_{2}$O$^+$ & OH$^+$ & SH & OH & CH & C\,{\small II} & O\,{\small I}\\
    \hline
        W3(OH) & -- & --~\,(3) & x~\,(3) & x & x & x~\,(6) & x & x\\
        W3~IRS5 & --~\,(1) & --~\,(3) & x\,~(3) & x & x & x~\,(6) & x~\,(8) & x\\
        NGC~7538~IRS1 & -- & -- & x & x & x & x & x & x\\
        DR21 &  --& --& --& --& -- & x & x~\,(8) & x \\
        IRAS~16060$-$5146 &  x$^{\prime}$\,(2) & x$^{\prime}$ & x$^{\prime}$ & x$^{\prime}$ & x$^{\prime}$\,(5) & x$^{\prime}$\,(7) & x$^{\prime}$ & x$^{\prime}$\,(5)\\
        IRAS~16164$-$5046 &  x$^{\prime}$\,(2) & x$^{\prime}$ & x$^{\prime}$ & x$^{\prime}$ & x$^{\prime}$\,(5) & --~\,(7) & -- & x$^{\prime}$\,(5)\\
        NGC~6334~I &  x$^{\prime}$ & x$^{\prime}$\,(3) & x$^{\prime}$\,(3) & x$^{\prime}$ & x$^{\prime}$ & x$^{\prime}$ & x$^{\prime}$ & x\\
        G10.47+0.03 & --~\,(2) & -- & -- & -- & --~\,(5) & --~\,(2) & x$^{\prime}$ &  x$^{\prime}$~(5)\\
        G29.96$-$0.02 &  x$^{\prime}$ & x$^{\prime}$\,(3) & x$^{\prime}$\,(3) & x$^{\prime}$\,(4) & x$^{\prime}$ & x$^{\prime}$ & x$^{\prime}$ & x$^{\prime}$\\
        G32.80+0.19 &  x$^{\prime}$ & x$^{\prime}$ & x$^{\prime}$ & x$^{\prime}$ & x$^{\prime}$ & x$^{\prime}$ & x$^{\prime}$ & x$^{\prime}$\\
        G45.07+0.13 & x$^{\prime}$ & x$^{\prime}$ & x$^{\prime}$ & x$^{\prime}$ & x$^{\prime}$ & x$^{\prime}$ & x$^{\prime}$ & x$^{\prime}$\\

         \hline
    \end{tabular}
    \end{center}
    
    \tablecomments{The columns demarcated using the x symbol denotes observations collected as a part of SOFIA's Cycle 8 northern deployment from K\"{o}ln-Bonn, Germany in February-March, 2021 while those marked with an x$^{\prime}$ correspond to data collected during the Cycle 9 southern deployment from Papeete, Tahiti in July-August 2021.\\
    Previous detections: (1)~unpublished data observed under the WISH survey using \textit{Herschel}/HIFI (Observation IDs: 1342201530 and 1342201531); (2)~\citet{Jacob2020Arhp};  (3)~\citet{Indriolo2015}; (4)~\citet{Neufeld2015S}; (5)~\citet{Wiesemeyer2016};  (6)~\citet{Wiesemeyer2018};
    (7)~\citet{Jacob2019};
    (8)~\citet{Gerin2015}.}
    
\end{table*}

\subsection{Ancillary data sets} \label{subsec:complementary_data}
In this section we briefly summarize the ancillary data collected using various ground-based facilities, which have proven to be useful in interpreting previous observations of the FIR and submillimeter transitions of hydrides. Since many of the molecular ions targeted under the HyGAL Legacy program originate in diffuse atomic clouds where the H$_2$ fraction is small, absorption-line measurements of the 21~cm H\,{\small I} line have proven essential for their interpretation and the determination of their abundances within such clouds \citep[see for example the use of such data as demonstrated in][]{Winkel2017}. Therefore, in order to exploit the diagnostic powers of these species and provide essential information for our modeling efforts, we have carried out observations of the \hi\ 21~cm line using the \textit{Karl G. Jansky} Very Large Array (JVLA) for the northern sources (Project ID: 20A-519, 21A-287; PI: M. R. Rugel) and archival Australia Telescope Compact Array (ATCA) observations for the southern sources, respectively. Simultaneously with the JVLA \hi\ observations we observed all four hyperfine-structure transitions of the OH ground state near 18~cm. These OH observations together with the 2.5~THz rotational transitions of OH observed with SOFIA, will provide a unique and valuable data set for this species. The reliable column densities determined from the SOFIA observations will aid in determining the excitation conditions of the former, which are usually not in thermal equilibrium. 

For the interpretation of SH, observations of other sulfur-bearing species such as CS, H$_2$S and SO obtained with the Instituto de Radioastronom\'ia Milim\'etrica (IRAM) 30~m telescope have proven very valuable \citep{Neufeld2015S}. Moreover the 3-mm wavelength band covering these species will also cover transitions of  other molecular gas tracers like HCO$^+$ and C$_{2}$H. Therefore, the SOFIA HyGAL observations will be analyzed alongside ancillary 2~mm and 3~mm wavelength observations carried out with the IRAM 30~m telescope (for the northern HyGAL sources) targeting primarily diatomic and triatomic species like SO, CS, C$_{2}$H, HCO$^+$, HOC$^+$, HCN, HNC and H$_2$S (Project ID: 003-20; PI: D. A. Neufeld). For the southern sources, we use archival data collected as a part of the ALMA Three-millimeter Observations of Massive Star-forming regions \citep[ATOMS;][]{Liu2020} survey (Project ID: 2019.1.00685.S; PI: Tie Liu). Detailed descriptions for each of the supplementary data sets discussed above and the results obtained, will be presented in forthcoming papers.


Absorption line studies of CO, in combination with \textit{Herschel} archival observations of [C~{\small I}] and [C~{\small II}] observations obtained as a part of the SOFIA HyGAL project, will allow us to better understand the C$^+$/C/CO (H~{\small I}-to-H$_2$) transition and the carbon budget (fraction of gas-phase carbon) in diffuse and translucent clouds. To reliably estimate CO column densities, we have utilized the excellent sensitivity provided by the NOrthern Extended Millimeter Array (NOEMA) to observe the $J =1-0$ lines of CO and its isotopologues; $^{13}$CO, C$^{17}$O and C$^{18}$O (Project IDs: S21AQ, W21AE; PI:W.-J., Kim). Interferometeric observations filters out extended emission, allowing the isolation of absorption features towards the background point source in a clean way. As discussed by \citet{Liszt1998} the combination of absorption and emission measurements provide the necessary information to probe the CO excitation and the CO column densities.
The value of CO absorption line measurements in probing CO in diffuse gas has previously been demonstrated by \citet{Liszt1998, Liszt2004, Liszt2019}.

\section{First results toward W3(OH), W3~IRS5, and NGC~7538~IRS1} \label{sec:first_results}
This section presents a first impression of the quality and analysis of the data collected as a part of the HyGAL survey, through examples toward three of the targeted sources in our sample - W3(OH), W3~IRS5, and NGC~7538~IRS1.

\subsection{Spectra and line-of-sight properties} \label{subsec:spec_los_props}

\begin{figure}
    \centering
    \includegraphics[width=0.49\textwidth]{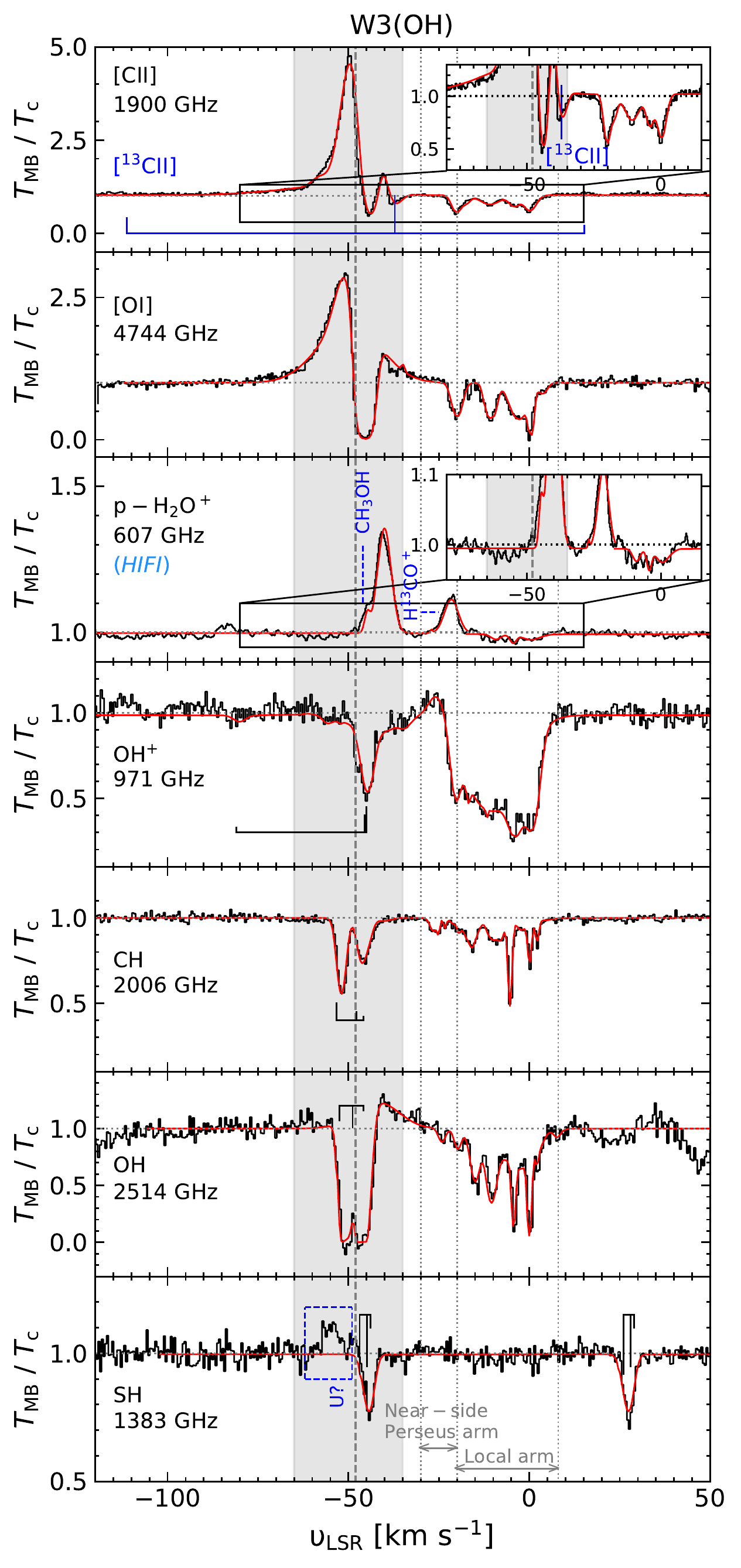}
    \caption{From top to bottom: Spectra in $T_{\rm MB}$ scales normalized with respect to the continuum temperature, $T_{\rm c}$, showing transitions of C$^+$, O, p-H$_{2}$O$^+$ (observed with \textit{Herschel}/HIFI and taken from \citet{Indriolo2015}), OH$^+$, CH, OH, and SH, respectively towards W3OH, in black, with the XCLASS fits in red. The inset in the top panel zooms-in on the C$^+$ absorption features over a range of $\upsilon_{\rm LSR}$ between $-80$ and 15~km~s$^{-1}$. The vertical gray dashed line and gray shaded regions mark the systemic velocity and the typical velocity dispersion of the source.
    The relative intensities of the hyperfine structure components of the OH$^+$, CH, and OH transitions are shown in black and $^{13}$C$^+$ in blue above/below their respective spectra. Contaminating emission features that are identified are labeled in blue and their contributions modeled and removed in all subsequent analysis, while those that remain unidentified are marked (U?).}
    \label{fig:w3oh_spec}
\end{figure}

Located in the outer Galaxy, W3(OH), W3~IRS5, and NGC~7538~IRS1 were observed as a part of SOFIA's Cycle 8 northern campaign from K\"{o}ln-Bonn, Germany. The data were collected during five flights on 2021 February 9, 10, 17, 18, and 19  with flight ids: 695, 696, 698, 699, and 700, respectively. The average leg time (including overheads) was 148~min per source for both the LFA+HFA and 4GREAT+HFA configurations (see Sect.~\ref{sec:observations} for a complete description of the observational setups used) with an average system temperature of 3370, 5540, 2750, 4240, and 13500~K for the LFA, HFA, 4G2, 4G3 and 4G4 receivers, respectively. At the time when these observations were carried out, the 4G1 channel was not functional and the ArH$^+$ $J=1-0$ and p-H$_{2}$O$^+$ $N_{K_{\rm a},K_{\rm c}} = 1_{1,0}-1_{0,1}$ lines were therefore not observed. Data products for all the other targeted lines were obtained for these three sources.  
Toward W3~IRS5, the ArH$^+$ $J=1-0$ and p-H$_{2}$O$^+$ $N_{K_{\rm a},K_{\rm c}} = 1_{1,0}-1_{0,1}$ lines had been collected as a part of the guaranteed time key program Water In Star-forming regions with \textit{Herschel} \citep[WISH;][]{vanDishoeck2011} (observation IDs: 1342201530 and 1342201531) and were retrieved from the \textit{Herschel} science archive\footnote{\href{https://archives.esac.esa.int/hsa/whsa/}{https://archives.esac.esa.int/hsa/whsa/}.}. Similarly, we present p-H$_{2}$O$^+$ data toward W3(OH) collected as a part of \textit{Herschel} open time programs with observation IDs: 1342268579-81 and published in \citet{Indriolo2015}. The resulting spectra of all observed species toward each of these sources are displayed in Figures~\ref{fig:w3oh_spec} to \ref{fig:ngc7538_spec}. 

The following paragraphs describe the individual line profiles and line-of-sight features toward W3(OH), W3~IRS5, and NGC~7538~IRS1,  while the properties and expected line-of-sight features (as determined from observations of other molecular species) for the remaining sources are given in Appendix~\ref{sec:los_source_properites}. Table~\ref{tab:SSB_continuum_noise} lists the single-sideband (SSB) continuum levels ($T_{\rm c}$) on main beam temperature scales toward this sub-sample of sources at each observed frequency, along with the root-mean-square (rms) noise in the corresponding spectra. 

The spectra observed along the line-of-sight towards hot-cores often show emission from a plethora of molecules, including complex organic species \citep[see, e.g.][]{Belloche2013}. Displaying a rich line spectrum at millimeter and submillimeter wavelengths, the emission from these species, many of which remain unidentified are often referred to as `weeds' as they contaminate the absorption profiles. In order to gauge the true depth of the absorption features we identify and model contributions from the contaminating species. The contamination observed in each spectral line profile, if any, toward each individual source is discussed in more detail in the following sections.
\begin{table*}
    \centering
        \caption{SSB continuum levels and rms noise of the spectra, and column density sensitivity.}
    \begin{tabular}{llccc}
    \hline \hline 
         Frequency &  & \multicolumn{3}{c}{Source}\\
         ~[GHz] &  & W3(OH) & W3~IRS5 & NGC~7538~IRS \\
         \hline
         971 & $T_{\rm c}$ [K] & 1.53 & 1.24 & 1.10\\
              & $T_{\rm rms}$ [K] & 0.04 & 0.08 & 0.14 \\ 
              & (d$N$(OH$^+$)/d$\upsilon$)$_{\rm rms}$  & 2.0$\times10^{10}$
              & 4.8$\times10^{10}$
              & 9.3$\times10^{10}$\\
         1383 & $T_{\rm c}$ & 2.81 & 2.55 & 1.96 \\
              & rms & 0.06  & 0.14 & 0.07 \\
              & (d$N$(SH)/d$\upsilon$)$_{\rm rms}$  & 1.6$\times10^{11}$ &
              4.0$\times10^{11}$ &
              0.6$\times10^{11}$\\
         1900 & $T_{\rm c}$ & 5.20 & 5.57 & 3.14\\
              & rms & 0.07  & 0.06 & 0.06\\
              & (d$N$(C$^+$)/d$\upsilon$)$_{\rm rms}$  & 1.0$\times10^{15}$ & 
              0.7$\times10^{15}$ & 
              1.2$\times10^{15}$\\
         2006 & $T_{\rm c}$ & 5.30 & 5.48 & 3.26 \\
              & rms & 0.04  & 0.04 & 0.06 \\
              & (d$N$(CH)/d$\upsilon$)$_{\rm rms}$  & 3.3$\times10^{10}$ &
              3.2$\times10^{10}$ &
              8.1$\times10^{10}$\\
         2514 & $T_{\rm c}$ & 6.33 & 7.85 & 4.42 \\
              & rms & 0.24  & 0.73 & 0.27\\ 
              & (d$N$(OH)/d$\upsilon$)$_{\rm rms}$  & 5.8$\times10^{10}$ &
              14.0$\times10^{10}$ & 
              9.3$\times10^{10}$\\
         4744 & $T_{\rm c}$ & 1.71 & 6.71 & 3.01\\
              & rms & 0.03  & 0.04 & 0.03\\
              & (d$N$(O)/d$\upsilon$)$_{\rm rms}$  & 6.3$\times10^{14}$ 
              & 2.1$\times10^{14}$ 
              & 3.6$\times10^{14}$\\
         \hline
    \end{tabular}
 \tablecomments{The rms noise level on the $T_{\text{MB}}$ scale is quoted for a spectral resolution of 0.5~km~s$^{-1}$.  (d$N$/d$\upsilon$)$_{\rm rms}$ represents the column density corresponding to $T_{\rm rms}$ in units of [cm$^{-2}$~/(km~s$^{-1}$)]. For those species with hyperfine-structure splitting, (d$N$/d$\upsilon$)$_{\rm rms}$ is computed using the spectroscopic parameters of the strongest hyperfine-structure transition.}
    \label{tab:SSB_continuum_noise}
\end{table*}
 
\subsubsection*{W3~IRS5 and W3(OH)}
The W3 molecular cloud complex is located in the second quadrant of the Galaxy in the Perseus arm and consists of two main high-mass star forming sub-regions, W3(OH) and W3 Main \citep[for an overview see for e.g.,][]{megeath2008low}, with a projected separation of $\sim$10~pc at the distance of the background sources ($\approx 16\rlap{.}^{\prime}$6 apart on sky). The archetypical ultracompact H{\small II} region W3(OH) \citep[e.g.,][]{Qin2016} is located at a distance of $\sim 2$~kpc \citep{Hachisuka2006, Xu2006} and thought to harbor a massive young star, while W3~Main contains several luminous infrared sources, of which the W3~IRS5 protocluster is the brightest \citep{Wang2013}. Moreover, the W3~IRS5 cluster system has a well known double IR source at the center of an embedded cluster of a few hundred low mass stars \citep{Megeath1996}. The velocity distribution towards W3(OH), shows two major components, of which one is attributed to the target itself and its envelope at velocities between $-70$ and $-35~$km~s$^{-1}$, and the other one to the local spiral arm between $-20$ and 7~km~s$^{-1}$. Despite having line-of-sight velocity components largely similar to that of W3(OH), the sight line toward W3~IRS5 with a systemic cloud velocity of $-39$~km~s$^{-1}$, shows relatively weaker absorption at velocities corresponding to the local spiral arm. This is particularly seen in the comparison of the CH and OH spectral line profiles toward both sources. While the deepest absorption features in the foreground material for CH and OH are observed at velocities between $-$7 and +5~km~s$^{-1}$ for W3(OH), those for W3~IRS5 lie between $-$32 and $-$12~km~s$^{-1}$, which is absorption attributed to the near side of the Perseus arm. The spatial separations for the foreground clouds (gas which is not locally associated with the source) traced by both sources range from 10~pc for clouds located near the sources to $<1$~pc further away from the background sources at positive velocities. Therefore, the HyGAL project also offers the possibility of studying spatial variations in the abundances of the species studied and in turn the gas properties traced along those sight lines in the survey that are spatially nearby.

\begin{figure}
    \centering
    \includegraphics[width=0.49\textwidth]{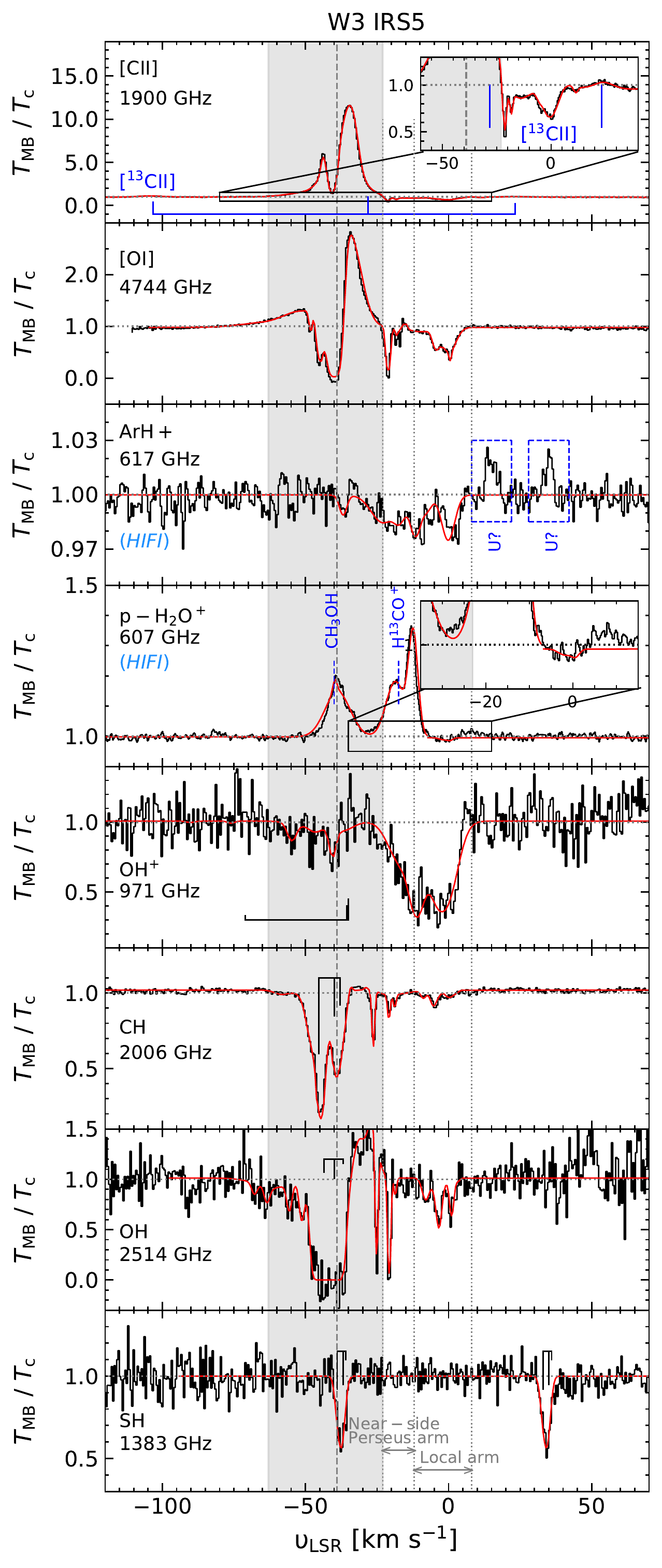}
    \caption{Same as Figure~\ref{fig:w3oh_spec} but toward W3~IRS5 with the addition of the ArH$^+$ $J=1-0$ spectrum observed using \textit{Herschel}/HIFI.}
    \label{fig:w3irs5_spec}
\end{figure}

 As discussed in Sect.~\ref{sec:observations} the p-H$_{2}$O$^+$ spectra observed toward both W3(OH) and W3~IRS5 are contaminated by high-lying H$^{13}$CO$^+$ and CH$_{3}$OH lines and contributions of these emission features are taken into account when modeling the p-H$_{2}$O$^+$ spectral line profiles. The ArH$^+$ spectrum along the line-of-sight toward W3~IRS5 also shows two emission features at $\upsilon_{\rm LSR}>10~$km~s$^{-1}$, both corresponding to the same transition and arising from the image band, which we were unable to identify. However, lying at velocities beyond the typical velocity range over which absorption features are observed (for example in comparison to the OH$^+$ spectrum displayed in Figure~\ref{fig:w3irs5_spec}), we do not expect any significant contributions from these lines. Similarly, the SH absorption feature at the systemic velocity of W3(OH) blends with an emission feature not seen in previous observations of SH toward other sight lines and showing only weak hints in the spectrum toward W3~IRS5 at $\upsilon_{\rm LSR} \sim -55~$km~s$^{-1}$. The non-detection of this feature toward other sight lines and its weaker presence toward W3~IRS5 at the same velocity suggests that this line arises from the image band. Covering only a narrow bandwidth and observed at only a single tuning frequency we are at this time unable to attribute this feature to any specific species. While this renders the true depth of the absorption line profile uncertain, the column densities that are subsequently derived toward the systemic velocity of the source (as will be discussed in more detail in Sect.~\ref{subsec:analysis}) only represent lower limits. 

\subsubsection*{NGC~7538~IRS1}
The vicinity of the H{\small II} region NGC~7538, located at a distance of 2.65~kpc \citep{Moscadelli2009} in the Perseus arm, is host to multiple infrared sources of which the brightest ones are NGC~7538~IRS~1, IRS~9 and South (S). The most massive of these, NGC~7538 IRS~1, an embedded O-type star, is a remarkably rich maser source at the heart of a young stellar cluster. Moreover, studies by \citet{Klaassen2009} on the dynamics of the warm molecular gas surrounding this hyper-compact H{\small II} region is indicative of rotation. With a systemic velocity of $-59~$km~s$^{-1}$, most of the absorption at $\upsilon_{\rm LSR}<-40~$km~s$^{-1}$ is associated with material from the background source, while the absorption features seen at $\upsilon_{\rm LSR}>-20~$km~s$^{-1}$ arise from foreground material in the local arm. The self-absorption features seen in the [C\,{\small II}] and [O\,{\small I}] spectra are offset from the systemic velocity of the IRS1 source near $\upsilon_{\rm LSR}\sim -55$~km~s$^{-1}$, in agreement with the complex kinematics of IRS1 discussed by \citet{Beuther2013}. 
\begin{figure}
    \centering
    \includegraphics[width=0.49\textwidth]{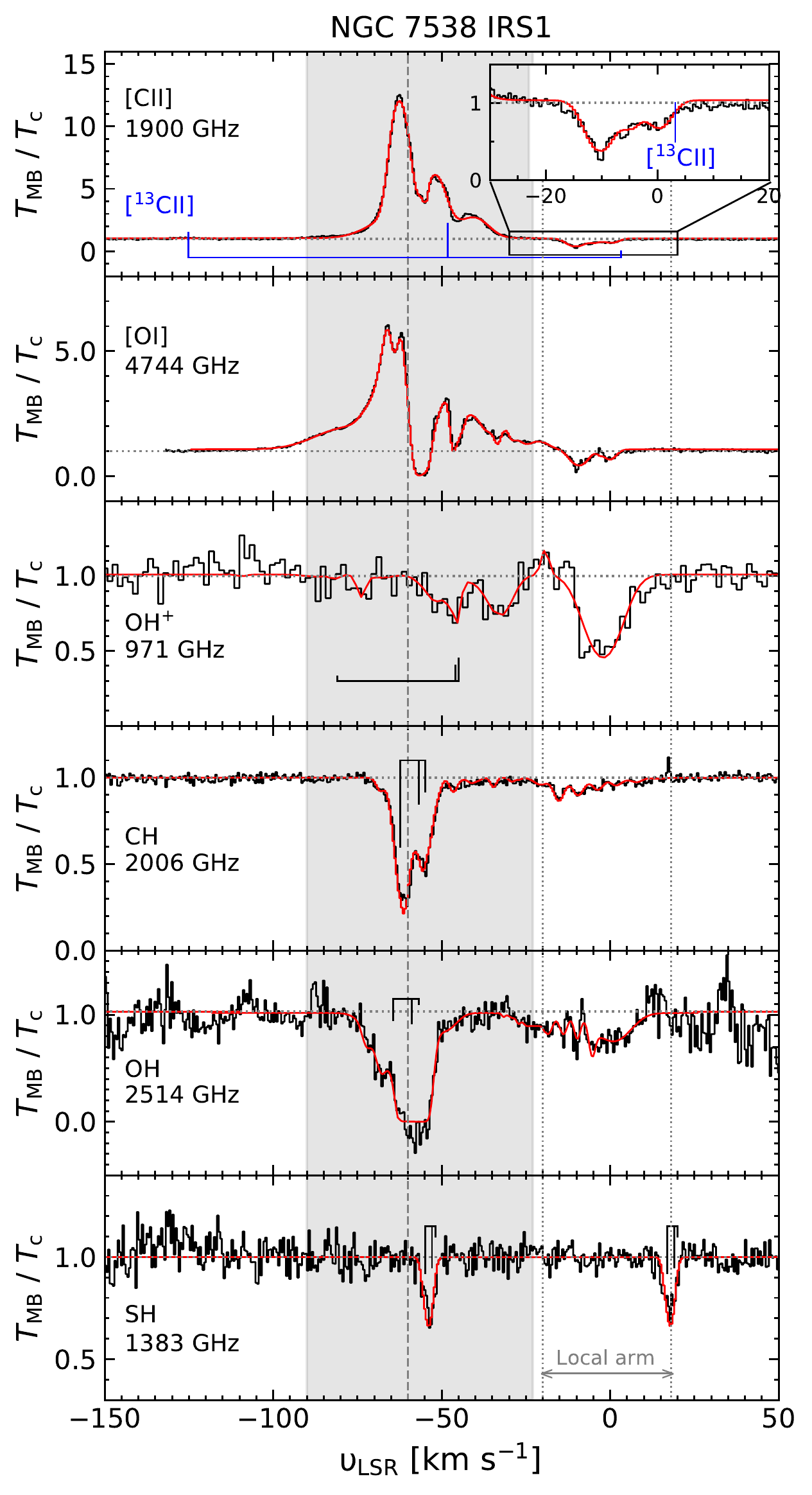}
    \caption{Same as Figure~\ref{fig:w3oh_spec} but toward NGC~7538~IRS1, with only the SOFIA/GREAT observations.}
    \label{fig:ngc7538_spec}
\end{figure}

\subsection{Analysis} \label{subsec:analysis}
The observed spectral line features are modeled using an enhanced version of the eXtended CASA Line Analysis Software Suite \citep[XCLASS\footnote{See, \href{https://xclass.astro.uni-koeln.de/}{https://xclass.astro.uni-koeln.de/} for further information.},][]{Moller2017}, which solves the 1D radiative transfer equation for an isothermal source assuming that the level populations are governed by Boltzmann statistics at a common temperature, $T_{\rm rot}$, such that the excitation temperature, $T_{\rm ex}$, is the same for all pairs of states. The spectral line profiles are fitted with multiple Gaussian components, taking into account the effects of optical depth on the line shape for a finite source size and dust attenuation using the automated fitting routine provided by MAGIX \citep{Moller2013}. From the resultant modeled optical depth profile, we determine the column densities per velocity interval, d$N$/d$\upsilon$, using
\begin{equation}
    \left( \frac{\text{d}N}{\text{d}\upsilon}\right)_{i} = \frac{8\pi\nu^3_{i} }{c^3 } \frac{Q(T_{\text{rot}})}{g_{\text{u}} A_{\text{u,l}}} \text{e}^{E_{\text{u}}/T_{\text{ex}}} \left[ \text{exp} \left(\frac{h\nu_{i}}{k_{\text{B}}T_{\text{ex}}}\right) - 1 \right]^{-1} \tau_{i}
    \label{eqn:col_dens}
\end{equation}
for each velocity channel, $i$. For a given hyperfine-structure splitting transition, the spectroscopic parameters $g_\text{u}$ (the upper level degeneracy), $E_\text{u}$ (the upper level energy) and $A_\text{u,l}$ (the Einstein A coefficient between any two given levels u and l) remain constant, except for the partition function, $Q$, which itself is a function of the rotation temperature, $T_{\text{rot}}$. 
We further assume that almost all the hydrides and hydride ions in this study occupy the ground rotational state where the excitation temperature, $T_{\text{ex}}$, is determined by the CMB radiation temperature, $T_{\text{CMB}}$, of 2.73~K. This is a valid assumption for the conditions that prevail in diffuse clouds where the gas densities, (${n(\text{H}) \leq  100~\text{cm}^{-3}}$), are sufficiently low that collisional excitation becomes unimportant. While it has been shown through the comparison of CH column densities derived
from its 532~GHz and 2007~GHz transitions that weak pumping by the ambient far-infrared radiation field can increase the excitation
temperature slightly above the CMB value, this effect only has a small impact on the derived column densities \citep{Wiesemeyer2018}. Changing the assumed $T_{\rm ex}$ from 3.1 to 10~K only results in a $\sim 10\%$ increase in the derived
column density. Similarly, by estimating the abundance of H$_2$O in its excited levels, \citet{Flagey2013} were able to constrain $T_{\rm ex}\approx 5~$K, for the ground state transitions of its ortho-spin state.
However, the assumption of a low excitation temperature, equivalent to the CMB value may not be valid for those features that arise from the dense envelopes of the molecular clouds in which collisions can dominate, resulting in higher excitation temperatures \citep{Emprechtinger2013}. Therefore, the column densities derived by integrating Eq.~\ref{eqn:col_dens} over velocity intervals associated with the molecular cloud components only represent lower limits to the column density values. 

The spectra of certain species (such as those of C\,{\small II} and O\,{\small I}, see the top panels of  Figures~\ref{fig:w3oh_spec} to \ref{fig:ngc7538_spec}), which have lower critical densities and higher optical depths than the hydride species, are more complex, containing both absorption and emission features near the envelope of the molecular cloud/hot-core. Here, the emission features corresponding to the core are modeled with higher excitation temperatures, $T_{\rm ex}\leq 120~$K, corresponding to gas temperatures typically found in the envelope material surrounding the central molecular cloud. Both the emission and  absorption features are fitted simultaneously by assuming different cloud layers. The former corresponds to a core component which probes the dense, hot cores associated with the studied molecular clouds, while the modeling of the latter is implemented using foreground layers which mimics the source envelopes and line-of-sight.

\begin{figure}
    \centering
    \includegraphics[width=0.48\textwidth]{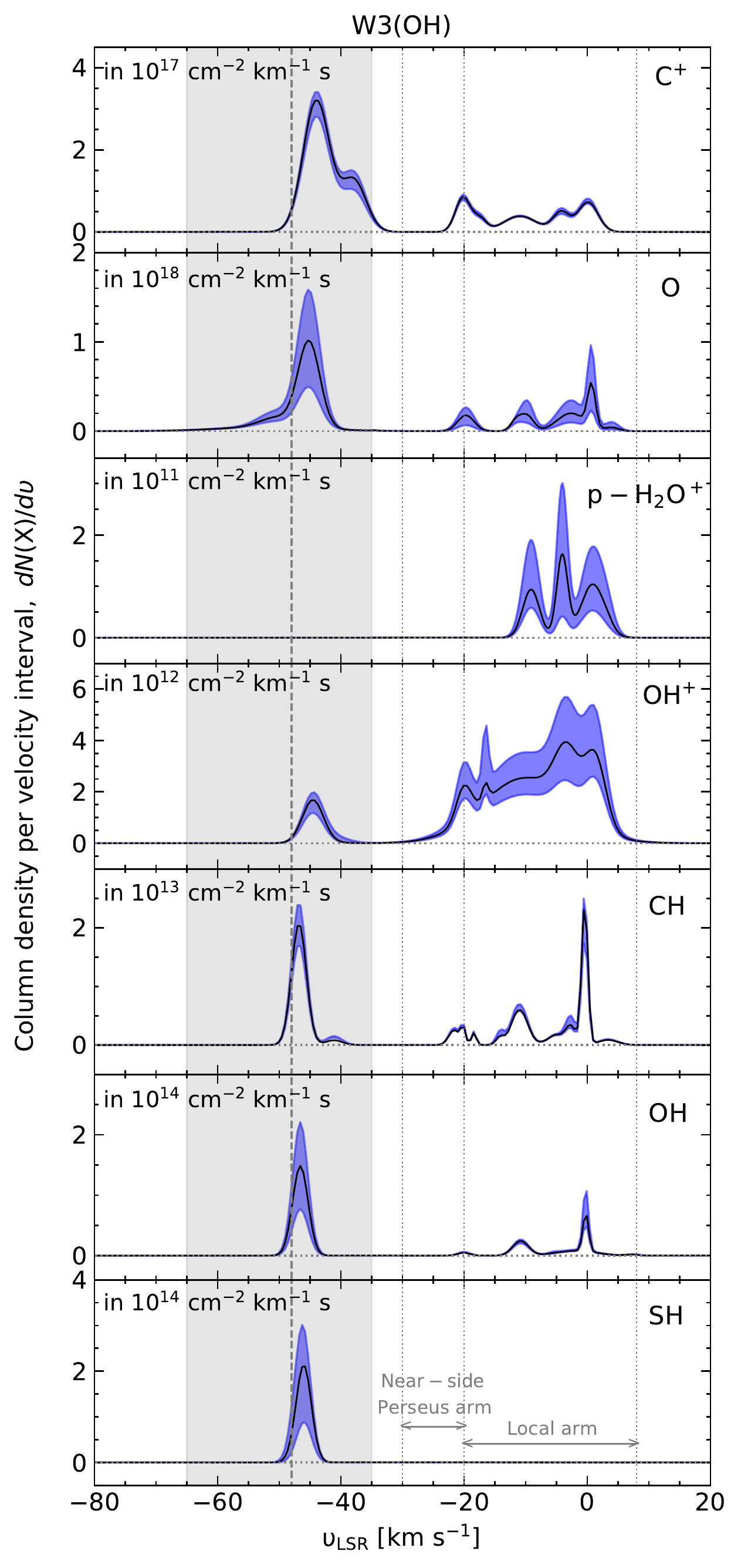}
    \caption{Top to bottom: Channel-wise column density profile (d$N$/d$\upsilon$) of C$^+$, O, p-H$_{2}$O$^+$, OH$^+$, CH, OH, and SH for the foreground absorbing clouds along the line-of-sight toward W3(OH). The blue shaded region represents the uncertainties (2$\sigma$ confidence interval) while the gray dashed line and shaded region mark the systemic velocity of the source and highlight the velocity dispersion of the source, respectively. }
    \label{fig:W3OH_coldens}
\end{figure}

\begin{figure}
    \centering
    \includegraphics[width=0.48\textwidth]{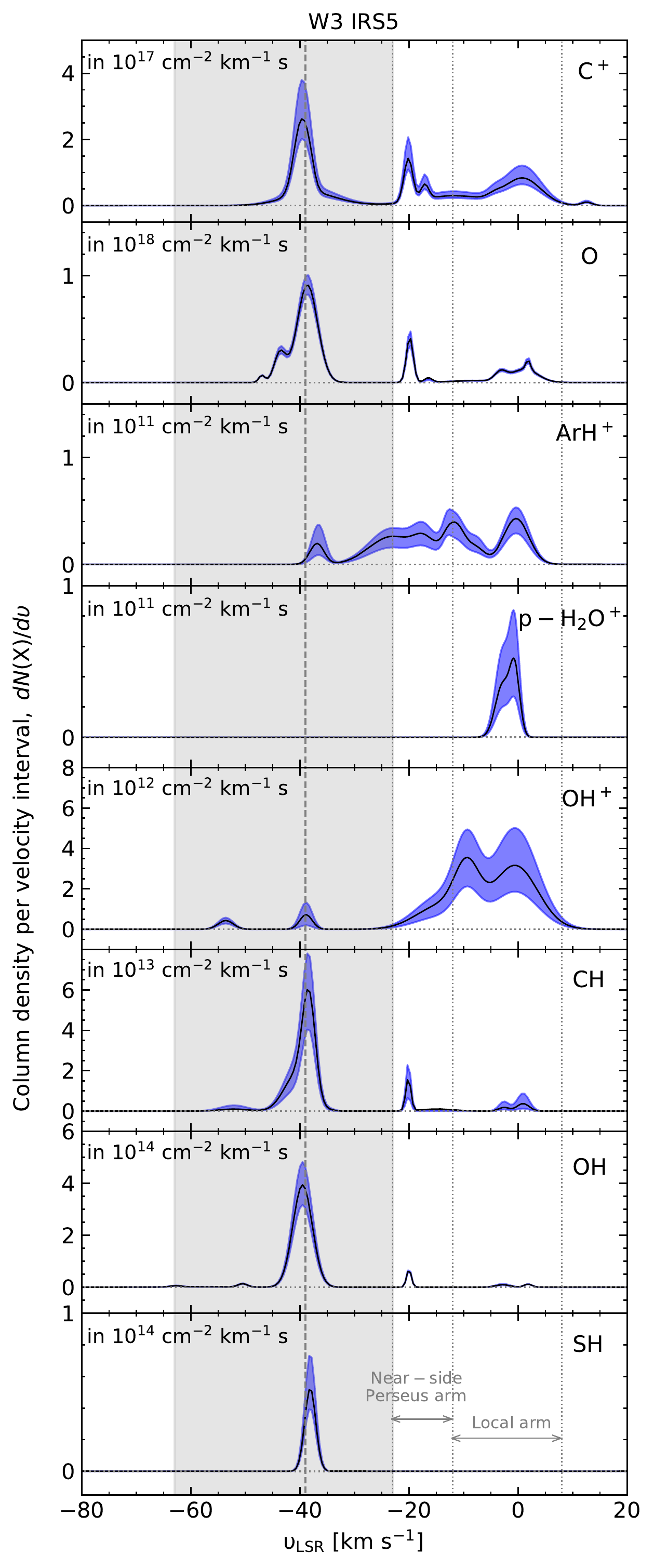}
    \caption{Same as Figure~\ref{fig:W3OH_coldens} but towards W3~IRS5 with the addition of the ArH$^+$ column density profile.}
    \label{fig:W3IRS5_coldens}
\end{figure}

\begin{figure}
    \centering
    \includegraphics[width=0.48\textwidth]{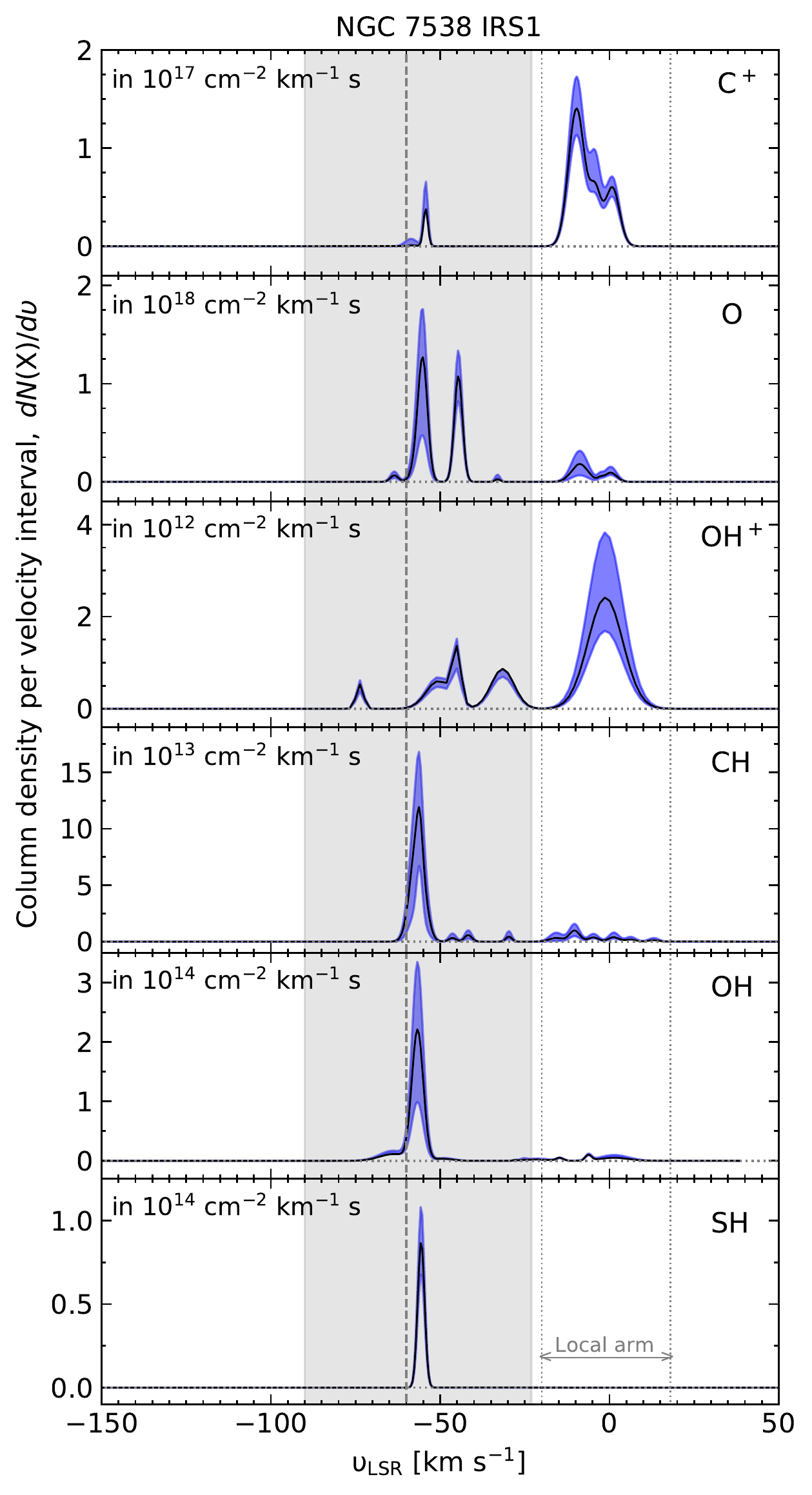}
    \caption{Same as Figure~\ref{fig:W3OH_coldens} but towards NGC~7538~IRS1 but without p-H$_{2}$O$^+$.}
    \label{fig:NGC7538_coldens}
\end{figure}

Contributions from the column densities subsequently determined for the emission features are removed from the total distribution revealing only the column density distribution of the underlying absorption from foreground material not associated with the source. The resulting column density profiles are displayed in Figures~\ref{fig:W3OH_coldens} to \ref{fig:NGC7538_coldens}. The total column densities estimated over specific velocity intervals, each of which broadly correspond to a spiral arm crossing are summarized in Table~\ref{tab:column_densities}. The tabulated column densities represent the total columns, which for those species with $\Lambda$-doublet splitting contain contributions from both sets of doublets.


The reliability of the column densities derived here are mainly limited by uncertainties present in the continuum level and not by the absolute temperature calibration. We compute the errors using Bayesian methods to sample a posterior distribution of the optical depths ($\propto {\rm ln}(T_{\rm l}/T_{\rm c}$)). We sampled a set of model spectra, each generated by iteratively adding a pseudo-random noise contribution to the absorption spectra, for each of the different species prior to running the XCLASS fits. The number of samplers is set by evaluating the posterior precision and variance while the standard deviation in the additive noise is fixed to be the same as that in the line-free part of each spectrum. The best fit curves generated by XCLASS for each spectra samples a point in the channel-wise column density distribution for any given source along any given sight line. The 5~\% and 95~\% percentiles obtained from the distributions are then used to quantify the 2$\sigma$ confidence intervals.


\begin{table*}
    \centering
    \caption{Synopsis of derived column densities and inferred results.}
    \begin{tabular}{l l cc ccccc }
    \hline \hline
         Source & $\upsilon_{\rm LSR}$ range & $N$(ArH$^+$) & $N$(p-H$_{2}$O$^+$) & $N$(OH$^+$) & $N$(OH) & $N$(SH)\tablenotemark{a} & $N$(CH) &   \\
         Designation & [km~s~$^{-1}$] & [$10^{12}$~cm$^{-2}$] & [$10^{12}$~cm$^{-2}$] & [$10^{13}$~cm$^{-2}$] & [$10^{14}$~cm$^{-2}$] & [$10^{13}$~cm$^{-2}$] & [$10^{13}$~cm$^{-2}$]   \\
         \hline
         W3(OH) & $-51$ to $-39$\tablenotemark{*} & \multicolumn{1}{c}{...} & \multicolumn{1}{c}{...} & $>0.73$ & $>4.67$&  $>6.70$ &$>6.48$\\
         & $-25$ to \phantom{0}$-8$ & \multicolumn{1}{c}{...} & 0.21$^{+0.25}_{-0.13}$& 3.03$^{+1.60}_{-1.10}$&  0.90$^{+0.12}_{-0.12}$& $<0.10$ & 3.14$^{+0.30}_{-0.30}$\\
         & \phantom{0}$-8$ to \phantom{0}+9 & \multicolumn{1}{c}{...} & 0.91$^{+0.80}_{-0.55}$ & 3.86$^{+2.30}_{-1.70}$ & 1.41$^{+0.57}_{-0.28}$ & $<0.10$ & 4.60$^{+0.70}_{-0.66}$\\
         W3~IRS5 & $-55$ to $-35$\tablenotemark{*} & $>0.01$ & \multicolumn{1}{c}{...}& $>0.06$ & $>9.02$& $>1.10$& $>9.60s$ \\
         & $-28$ to \phantom{0}$-8$ &\phantom{0}0.51$^{+0.16}_
         {-0.19}$ & \multicolumn{1}{c}{...} & 2.22$^{+1.43}_{-0.97}$ & 0.75$^{+0.05}_{-0.05}$ & $<0.09$& 2.67$^{+1.22}_{-1.22}$\\
         & \phantom{0}$-8$ to $+10$ & \phantom{0}0.27$^{+0.08}_{-0.08}$ & 0.21$^{+0.22}_{-0.14}$ &3.48$^{+2.00}_{-1.64}$ & 0.51$^{+0.10}_{-0.07}$& $<0.10$& 1.50$^{+1.40}_{-1.00}$\\
         NGC~7538~IRS1 & $-65$ to $-39$\tablenotemark{*} & \multicolumn{1}{c}{...} & \multicolumn{1}{c}{...}&  $>0.93$ & $>10.44$ & $>22.00$ & $>55.68$\\
         & $-39$ to $-31$ & \multicolumn{1}{c}{...}& \multicolumn{1}{c}{...}& 0.35$^{+0.04}_{-0.03}$ & \multicolumn{1}{c}{...}& $<0.05$& 0.02$^{+0.02}_{-0.01}$ \\
         & $-31$ to $-20$ &  \multicolumn{1}{c}{...}& \multicolumn{1}{c}{...}& 0.28$^{+0.05}_{-0.05}$ &0.14$^{+0.10}_{-0.02}$ &$<0.07$ & 0.83$^{+0.60}_{-0.40}$ \\
         & $-17$ to \phantom{0}$-3$ & \multicolumn{1}{c}{...}& \multicolumn{1}{c}{...}& 1.01$^{+0.40}_{-0.31}$ &0.43$^{+0.22}_{-0.16}$ &$<0.13$ & 5.96$^{+1.60}_{-1.60}$\\
         & \phantom{0}$-3$ to +18 & \multicolumn{1}{c}{...}& \multicolumn{1}{c}{...}& 1.86$^{+0.62}_{-0.54}$ &0.45$^{+0.08}_{-0.05}$& $<0.09$&  2.82$^{+0.37}_{-0.26}$\\
         \hline 
    \end{tabular}
    \label{tab:column_densities}
\end{table*}
\begin{table*}[]
    \setcounter{table}{4}
    \caption{Continued}
    \begin{tabular}{l l cc ccccc }
    \hline \hline
         Source & $\upsilon_{\rm LSR}$ range & $N$(H)\tablenotemark{b} & $N$(H$_{2}$)\tablenotemark{c} & $N$(C$^{+}$) & $N$(O) \\
         Designation & [km~s~$^{-1}$] & [10$^{21}$~cm$^{-2}$] & [10$^{21}$~cm$^{-2}$] & [10$^{17}~$cm$^{-2}$] & [10$^{18}$~cm$^{-2}$] &    \\
         \hline
         W3(OH) & $-51$ to $-39$\tablenotemark{*} & 1.30$^{+0.02}_{-0.02}$ & $>1.85$ & $>23.00$ & $>5.57$\\
         & $-25$ to \phantom{0}$-8$ & 1.39$^{+0.03}_{-0.03}$ & 0.90$^{+0.54}_{-0.37}$ & 5.51$^{+0.40}_{-0.40}$ & 1.26$^{+0.68}_{-0.63}$ \\
         & \phantom{0}$-8$ to \phantom{0}$+9$ & 0.95$^{+0.03}_{-0.03}$ & 1.31$^{+0.81}_{-0.55}$ & 4.80$^{+0.57}_{-0.42}$ & 2.00$^{+1.56}_{-1.08}$\\
         W3~IRS5 & $-55$ to $-35$\tablenotemark{*} & $>2.60$ & $>2.74$ & $>5.96$ &  $>2.21$ & \\
         & $-28$ to \phantom{0}$-8$ & 1.20$^{+0.26}_{-0.26}$ & 0.76$^{+0.57}_{-0.46}$ & 6.75$^{+2.43}_{-2.04}$ & 0.83$^{+0.46}_{-0.34}$\\
         & \phantom{0}$-8$ to $+10$ & 0.97$^{+0.21}_{-0.21}$ & 0.42$^{+0.46}_{-0.32}$ & 8.00$^{+3.06}_{-2.40}$ &1.15$^{+0.63}_{-0.52}$\\
         NGC~7538~IRS1 & $-65$ to $-39$\tablenotemark{*}  & \multicolumn{1}{c}{...} & $>16.00$ & $>0.69$& $>8.24$\\
         & $-39$ to $-31$ & \multicolumn{1}{c}{...}& 0.01$^{+0.01}_{-0.01}$ & \multicolumn{1}{c}{...}& 0.05$^{+0.02}_{-0.02}$\\
         & $-31$ to $-20$ & \multicolumn{1}{c}{...}& 0.23$^{+0.21}_{-0.14}$ & \multicolumn{1}{c}{...}& \multicolumn{1}{c}{...}\\
         & $-17$ to \phantom{0}$-3$ & 2.96$^{+0.03}_{-0.03}$ & 1.70$^{+1.12}_{-0.82}$&  9.80$^{+2.70}_{-2.00}$& 1.01$^{+0.82}_{-0.70}$ \\
         & \phantom{0}$-3$ to +18 & 1.99$^{+0.05}_{-0.05}$& 0.81$^{+0.50}_{-0.33}$ & 3.33$^{+0.71}_{-0.54}$ & 0.42$^{+0.27}_{-0.11}$\\
         \hline 
    \end{tabular}
    \label{tab:column_densities_part2}
    \tablecomments
    {\tablenotetext{*}{Indicates the velocity intervals corresponding to the molecular cloud.}\tablenotetext{a}{Represents the 3$\sigma$ upper limits on $N$(SH) obtained toward line-of-sight material not associated with the background source for which we do not detect SH.}\tablenotetext{b}{The H\,{\small I} data toward W3(OH) and W3~IRS5 are extracted from the Canadian Galactic Plane Survey \citep[CGPS,][]{Taylor2003} and that toward NGC~7538~IRS1 is taken from \citet{Lebron2001}, respectively.\tablenotetext{c}{$N(\text{H}_{2})$ values were derived using $N(\text{CH})$ following the relationship determined by \citet{Sheffer2008}, $N$(CH)/$N$(H$_{2}$)= 3.5$^{+2.1}_{-1.4}\times10^{-8}$. }   }}
\end{table*}

\section{Discussion} \label{sec:discussion}
A full statistical investigation of the diffuse gas properties traced using hydrides (as outlined in Sect.~\ref{sec:goals}) will require a combination of the complete HyGAL SOFIA data set with the ancillary IRAM and JVLA observations. The goal of the discussion section in the present paper is more modest: here we merely illustrate the analysis techniques that can be used to explore the scientific questions addressed by the HyGAL project. In subsections \ref{subsec:coldensanalysis} and \ref{subsec:los_ratios} we discuss the column density ratios implied by our absorption line observations and how they vary. Although the goals and analysis of the HyGAL survey concentrate on the foreground diffuse clouds along the sight lines toward bright sources, the SOFIA HyGAL data set provides a wealth of information on the background sources as well. We demonstrate this in subsection \ref{subsec:12c13c_ratio} through the analysis of the complex line profiles of the 158~$\mu$m [C\,{\small II}] emission, using constraints from the [$^{13}$C\,{\small II}] line also covered in the same sideband.

\subsection{Column density ratios}\label{subsec:coldensanalysis}
\begin{figure*}
    \centering
    \includegraphics[width=0.98\textwidth]{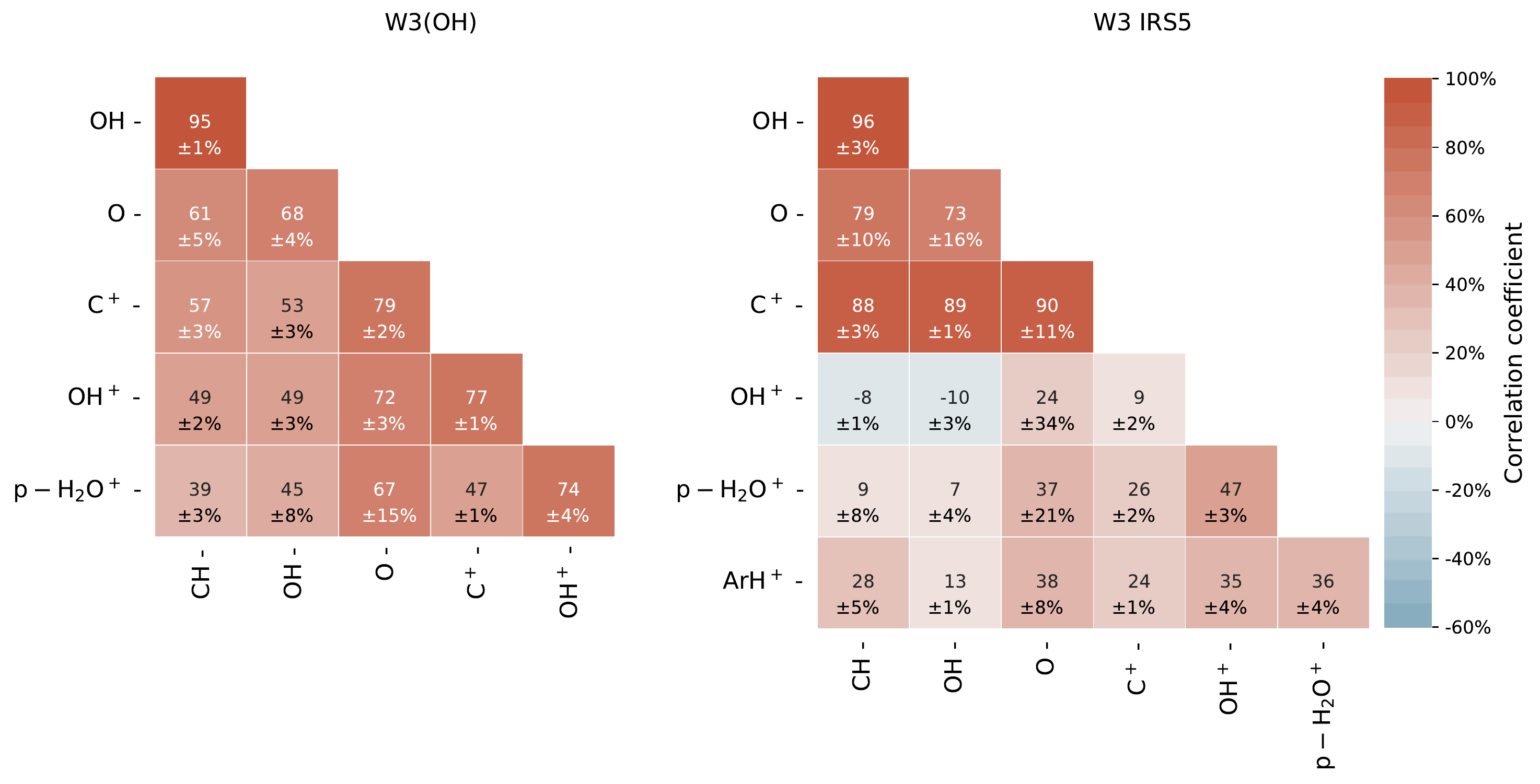}
    \caption{Pearson product-moment correlation coefficients as computed using Eq.~\ref{eqn:Pearsons_correlation_coeff} (represented here as percentages) between the column densities determined toward only those features which are not associated with the background source, for W3(OH) (left) and W3~IRS5 (right).  }
    \label{fig:all_cross_correlation}
\end{figure*}

\begin{figure}
    \includegraphics[width=0.5\textwidth]{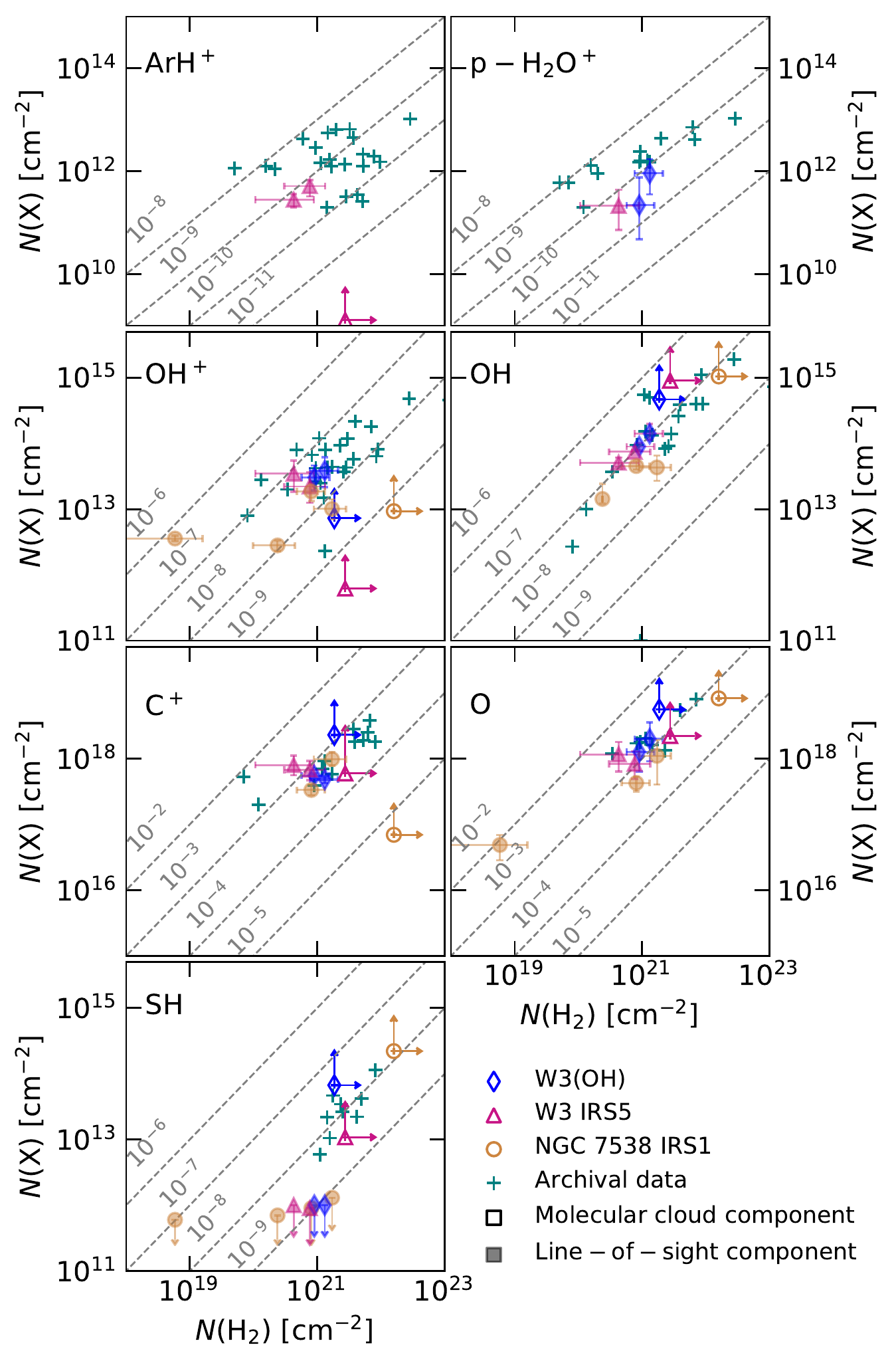}
        \caption{Comparison of column densities of the key HyGAL species with $N$(H$_{2}$) determined using $N$(CH)/$N$(H$_{2}$) = 3.5$\times10^{-8}$ \citet{Sheffer2008} and integrated over the velocity intervals discussed in Table~\ref{tab:column_densities}. The data points toward W3(OH), W3~IRS5, and NGC~7538~IRS1 are plotted using blue diamonds, pink triangles, and dark orange circles, respectively where the filled and unfilled markers demarcate the column densities determined toward line-of-sight components (features that are not associated with the background molecular cloud) and molecular cloud components as discussed in Table~\ref{tab:column_densities}. The dashed gray lines represent lines of constant abundances $N$(X)/$N$(H$_{2}$) for each studied species X as labelled on the top left-hand corner of each panel. The data represented using teal crosses present results from previous observations as discussed in the text.}
    \label{fig:Correlation_plots_H2}
\end{figure}

\begin{figure}
    \includegraphics[width=0.5\textwidth]{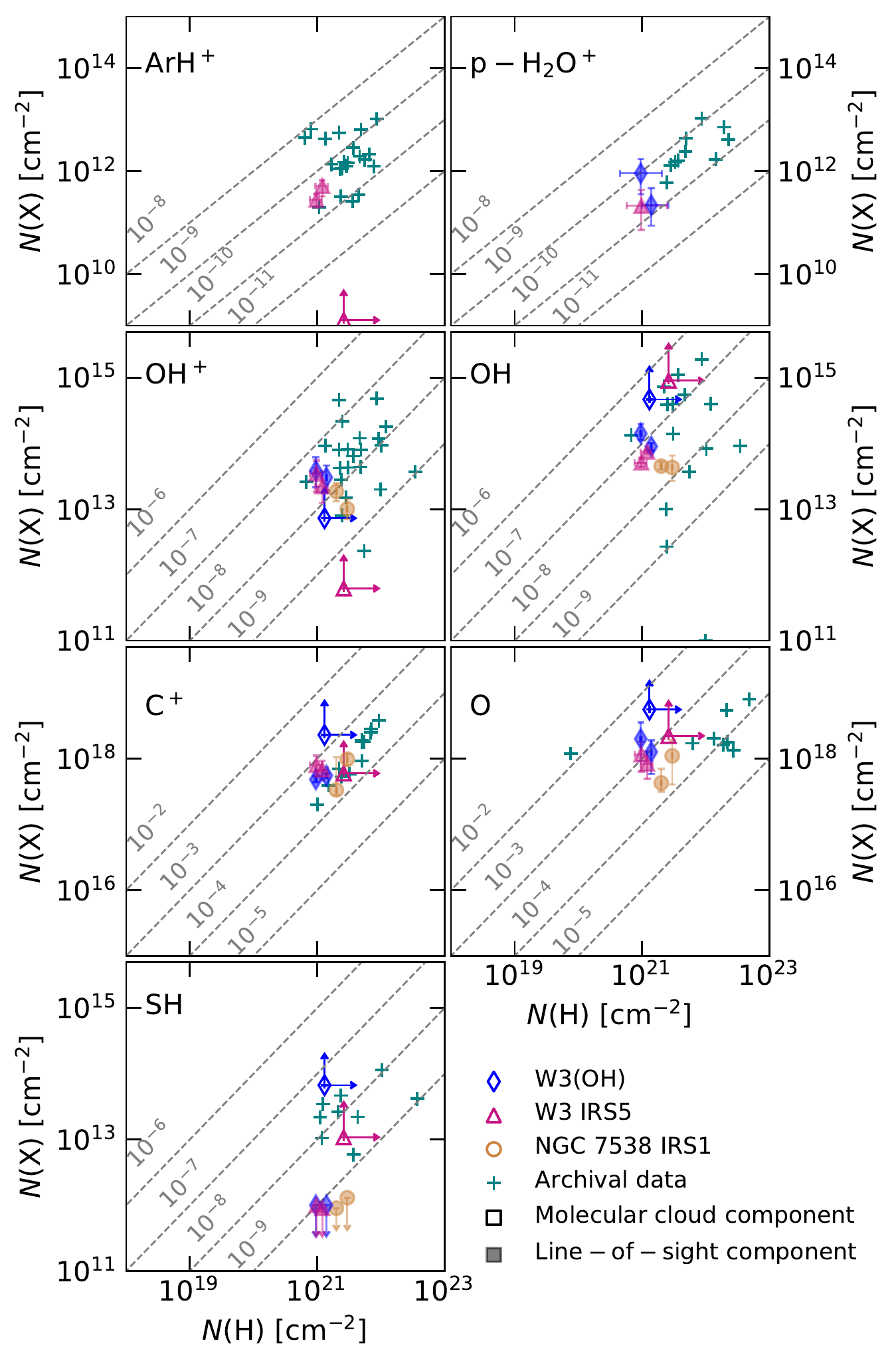}
    \caption{Same as Figure~\ref{fig:Correlation_plots_H2} but a comparison of column densities with respect to $N{\rm (H)}$.}
    \label{fig:Correlation_plots_H}
    \end{figure}
    
\begin{figure}
    \includegraphics[width=0.5\textwidth]{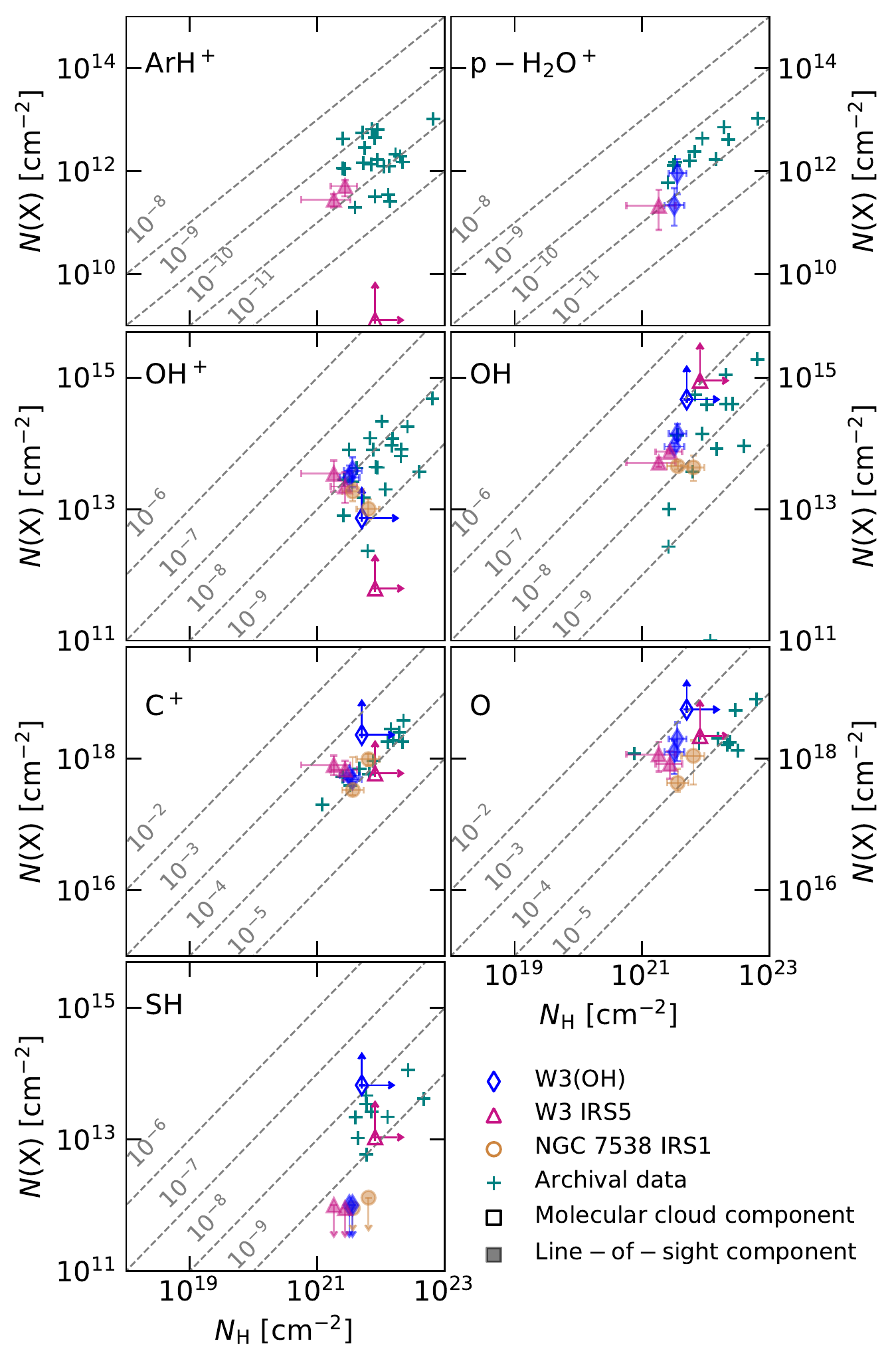}
    \caption{Same as Figure~\ref{fig:Correlation_plots_H2} but a comparison of column densities with respect to $N_{\rm H}$.}
    \label{fig:Correlation_plots_Htot}
    \end{figure}

From section~\ref{subsec:analysis} it is clear that the line profiles of some of the studied species are much more similar in shape across any given sight line than others. For example, the spectral line profiles of the diffuse molecular gas tracers in our survey, CH and OH, show similar absorption line features across the range of velocities probed. Their spectra reveal overall narrower features ($\sim 0.5~$km~s$^{-1}$) in comparison to the broader and more continuous profiles traced by the diffuse atomic gas tracers studied like OH$^+$. Moreover, the trends in absorption line strengths observed --  and in turn the abundances derived -- across a sight line differ from one hydride to another. Therefore, in this section we present a quantitative discussion of the observed line-of-sight properties using the column densities derived as discussed in Sect.~\ref{sec:first_results}. This is best illustrated in Figure~\ref{fig:all_cross_correlation}, which presents cross-correlations between the column densities of different pairs of species studied. The correlations presented are computed using the Pearson product moment correlation coefficient, $r$, which describes the strength of the linear relationship between any given pair of spectral lines by using the standard deviation of each data set and the covariance between them. The correlation coefficient between any two species $X$ and $Y$ is computed as follows,
\begin{equation}
    r = \frac{\int {\rm d}\upsilon \left( \frac{{\rm d}N(X)}{{\rm d}\upsilon} - \langle\frac{{\rm d}{N}(X)}{{\rm d}\upsilon}\rangle \right)\left( \frac{{\rm d}N(Y)}{{\rm d}\upsilon} - \langle\frac{{\rm d}{N}(Y)}{{\rm d}\upsilon}\rangle \right)}{\sqrt{\int {\rm d}\upsilon  \left( \frac{{\rm d}N(X)}{{\rm d}\upsilon} - \langle\frac{{\rm d}{N}(X)}{{\rm d}\upsilon}\rangle \right)^{2}} \sqrt{\int {\rm d}\upsilon \left( \frac{{\rm d}N(Y)}{{\rm d}\upsilon} - \langle\frac{{\rm d}{N}(Y)}{{\rm d}\upsilon}\rangle \right)^2}}
    \label{eqn:Pearsons_correlation_coeff}
\end{equation}
where, ${\rm d}N/{\rm d}\upsilon$ and $\langle {\rm d}\bar{N}/{\rm d}\upsilon \rangle$ represent the column densities computed per velocity channel and the average of the column densities computed over all relevant velocity channels.

In Figures~\ref{fig:Correlation_plots_H2} to \ref{fig:Correlation_plots_Htot}, we compare the column densities of all the studied species with that of molecular hydrogen, atomic hydrogen and the total gas column, $N_{\rm H} = N({\rm H}) + 2\times N({\rm H}_{2})$ derived over specific velocity intervals, respectively. For comparison, we also include the column density results for the same species obtained from previous \textit{Herschel}/HIFI and SOFIA/GREAT observations carried out toward other star-forming regions in the Galaxy. The data used in this comparison is primarily taken from \citet{Schilke2014} and \citet{Jacob2020Arhp} for ArH$^+$, \citet{Indriolo2015} and \citet{Jacob2020Arhp} for p-H$_{2}$O$^+$, \citet{Indriolo2015} for OH$^+$, \citet{Wiesemeyer2016} for OH and atomic oxygen, \citet{Gerin2015} for C$^+$, and \citet{Neufeld2015S} for SH.

Widely established and used as a tracer of CO-dark molecular gas \citep{Federman1982, Sheffer2008, Gerin2010CH, Wiesemeyer2018, Jacob2019}, CH is used as a proxy for H$_{2}$ in our analysis, using the relationship determined by \citet{Sheffer2008}, $N$(CH)/$N$(H$_{2}$)= 3.5$^{+2.1}_{-1.4}\times10^{-8}$. Since we determine $N$(H$_{2}$) using the $N$(CH) values derived, we do not include CH in our abundance analysis. 

Limited to a small sample of data points, a clear correlation is only discerned for the oxygen-bearing species. As discussed in Sect.~\ref{subsec:crir} the formation of OH$^+$ is initiated by cosmic rays, involving reactions with atomic and molecular hydrogen while the formation of OH greatly relies on the H$_{2}$ fraction. This is well reflected in the column density comparisons presented in Figures~\ref{fig:Correlation_plots_H2} to \ref{fig:Correlation_plots_Htot}. While both OH and atomic oxygen show positive correlations with increasing $N$(H$_{2}$) or molecular content, OH$^+$ shows no clear correlation with a larger scatter. The tight correlation observed between OH and CH (and in turn H$_{2}$) has also been explored at UV wavelengths by \citet{Weselak2009} and \citet{Mookerjea2016}, who derive a $N$(OH)/$N$(CH) ratio of ${\sim 2.6}$, equivalent to a $N$(OH)/$N$(H$_{2}$) ratio of ${\sim 10^{-7}}$ (as shown in Fig.~\ref{fig:Correlation_plots_H2} and Fig.~8 of \citet{Jacob2019}). At the Galactocentric distances of these sources the average atomic oxygen abundance derived with respect to $N_{\rm H}$ of $\sim4\times10^{-4}$ is consistent with what is expected when using the Galactic [O\,{\small I}] gradient as estimated by \citet{Korotin2014}, who determined a slope of $-$0.058~dex~kpc$^{-1}$ using observations of the near-infrared O\,{\small I} triplet toward a large sample of Cepheids.  

While no clear trend is discernible for the atomic gas tracers, ArH$^+$ and p-H$_{2}$O$^+$ (even with respect to N(H), see Fig.~\ref{fig:Correlation_plots_H}) where the usefulness of the latter is compromised by contaminating features, the abundances of the line-of-sight components are consistent with what was derived in \citet{Indriolo2015, Bialy2019} and \citet{Jacob2020Arhp}, of a few times 10$^{-10}$. The lack of a clear correlation with $N$(H) also suggests that only a small fraction of the H\,{\small I} gas traces the same cloud population as that traced by ArH$^+$ and p-H$_{2}$O$^+$.

For SH, the derived abundances roughly span over an order of magnitude between 4$\times10^{-9}$ and $3.5\times 10^{-8}$, consistent with the range of abundances derived in \citet{Neufeld2015S}. However, unlike the SH spectra observed toward the sight lines studied in \citet{Neufeld2015S}, the SH spectra presented here do not show any foreground absorption from line-of-sight clouds which are not associated with the sources themselves. The upper limits on the SH column densities we derive at a 3$\sigma$ rms level for these features vary between $\sim 1\times 10^{12}$ and 3.8$\times 10^{12}$~cm$^{-2}$, which correspond to abundance limits between 5.1$\times10^{-10}$ and 5.6$\times10^{-9}$ relative to H$_{2}$.

We note that some of the upper limits derived here are smaller than the range of SH abundances derived for diffuse clouds along the lines-of-sight toward other star-forming regions by \citet{Neufeld2015S}. This suggests that the non-detection of SH is not solely an issue of sensitivity, but also that the abundance of SH along the sight lines presented is lower than those measured previously for foreground diffuse clouds. As in the case of SH, no line-of-sight absorption is seen in the spectra of related sulfur bearing species such as CS\footnote{Except for W3(OH), along the line-of-sight toward which we observe weak absorption against the local arm.}, H$_{2}$S and SO observed using the IRAM 30~m telescope discussed in Sect.~\ref{subsec:complementary_data} (the spectra of which will be presented in a forthcoming paper); neither is it seen in the SH$^+$ spectra taken from \textit{Herschel} science archives toward W3(OH), W3~IRS5, and NGC~7538~IRS1. However, foreground absorption from clouds not associated with the background sources are observed in the \textit{Herschel} spectra of CH$^+$, which like SH$^+$ traces turbulent dissipation regions. The presence of CH$^+$ and absence of SH$^+$ (and by extension SH) in these clouds can be explained by the large differences in their endothermicities ($\Delta E_{({\rm CH}^+)}/k$ = 4640~K and $\Delta E_{({\rm SH}^+)}/k$ = 9860~K) of the ion-molecular reactions that are responsible for their formation. This, in addition to the lower effective temperatures and velocity drifts required, favors the production of CH$^+$ over SH$^+$ as discussed in \citet{Godard2014}. Therefore, in general the lack of foreground absorption in sulfur-bearing species toward these sources might be because their short sight lines do not probe sufficiently shocked or turbulence dissipated environments necessary for the formation of these molecules.   

Since the molecular gas column traced along the sight lines discussed is greater than or equivalent to the atomic gas column, the overall trends observed when comparing the column densities of the different species studied with that of molecular hydrogen and the total gas column, are consistent with one another (see Figures~\ref{fig:Correlation_plots_H2} and \ref{fig:Correlation_plots_Htot}). The main difference arises from the fact that there are fewer atomic gas components measured toward NGC~7538~IRS1. This is because the H\,{\small I} spectrum toward NGC~7538~IRS1 is extracted from a neighboring region G111.61+0.37 \citep{Lebron2001} and therefore we can only assume that the foreground material along the lines-of-sight toward both clouds is the same. Eliminating such uncertainties the auxiliary data collected using the JVLA under the umbrella of the HyGAL program (see Sect.~\ref{subsec:complementary_data}) will greatly aid in the analysis and interpretation of the targeted hydrides and other small interstellar molecules.

\subsection{Variations along the sight line}\label{subsec:los_ratios}
\begin{figure*}[t]
    \centering
    \includegraphics[width=0.85\textwidth]{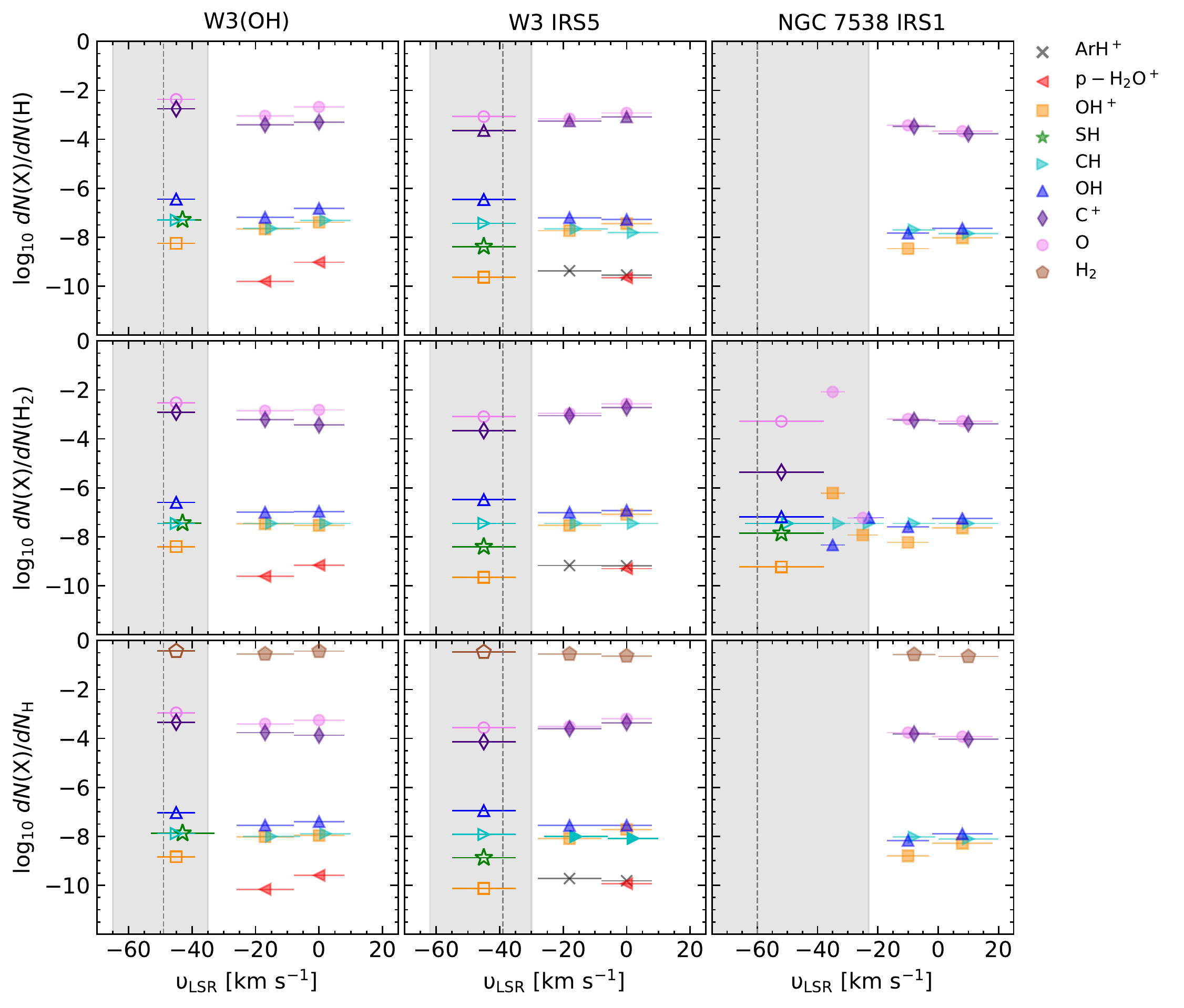}
    \caption{From top to bottom: Inferred $N$(X)/$N$(H), $N$(X)/$N$(H$_{2}$) and $N$(X)/$N_{\rm H}$ ratios over the adopted velocity intervals for W3(OH), W3~IRS5 and NGC~7538~IRS1 from left to right, with the different species denoted using different markers and colors as displayed in the legend. The filled and unfilled markers demarcate the column densities  determined toward line-of-sight and molecular cloud features, respectively. The gray dashed lines and shaded regions indicate the systemic velocity and velocity dispersion of the source, respectively, where the column densities determined only form lower limits. Markers lying close to each other are shifted by +2~km~s$^{-1}$ for clarity. Covering a large range of abundances, we do not display the corresponding errors in this plot as they coincide with the length scales of the markers.}
    \label{fig:column_density_ratio}
\end{figure*}

\begin{table*}
    \centering
        \caption{Range of abundances derived in the foreground absorbing clouds.}
    \begin{tabular}{l ccc}
         \hline \hline 
         Species & $N$(X)/$N$(H) & $N$(X)/$N$(H$_{2}$) & $N$(X)/$N_{\rm H}$ \\
         
        \hline
         ArH$^+$ & $6.73-8.42\times10^{-10}$ & $2.13-5.57\times10^{-10}$ & $0.92-2.10\times10^{-10}$ \\
         p-H$_{2}$O$^+$& $0.15-1.00\times10^{-9}$ & $1.00-3.16\times10^{-10}$ & $0.37-1.36\times10^{-10}$\\
         OH$^+$ & $0.34-4.06\times10^{-8}$ &  $0.60-3.55\times10^{-8}$ & $0.16-1.88\times10^{-8}$\\ 
         OH & $0.14-1.48\times10^{-7}$ & $2.53-6.01\times10^{-8}$&$0.67-2.02\times10^{-8}$\\
         C$^+$ & $1.67-8.23\times10^{-4}$ &  $1.58-8.14\times10^{-4}$&  $0.68-2.73\times10^{-4}$ \\
         O & $0.21-2.10\times10^{-3}$ & $0.47-8.31\times10^{-3}$ & $0.12-0.39\times10^{-3}$\\
         \hline
    \end{tabular}

    \label{tab:abundance_ratios}
\end{table*}

Figure~\ref{fig:column_density_ratio} displays the average column density ratios of the targeted species relative to atomic hydrogen, molecular hydrogen and the total gas column computed over the velocity intervals discussed in Table~\ref{tab:column_densities}. While the ratios computed for the velocity intervals associated with the background source are highly uncertain, the column density ratios even among line-of-sight clouds greatly deviate from one another. The range of abundances derived for each species toward the foreground material is tabulated in Table~\ref{tab:abundance_ratios}. The atomic gas tracer OH$^+$, unlike the molecular gas tracers CH and OH, displays the most stark variation in the column density ratios inferred toward the molecular cloud and line-of-sight velocities.\\ 

\begin{figure*}
    
    \includegraphics[width=1\textwidth]{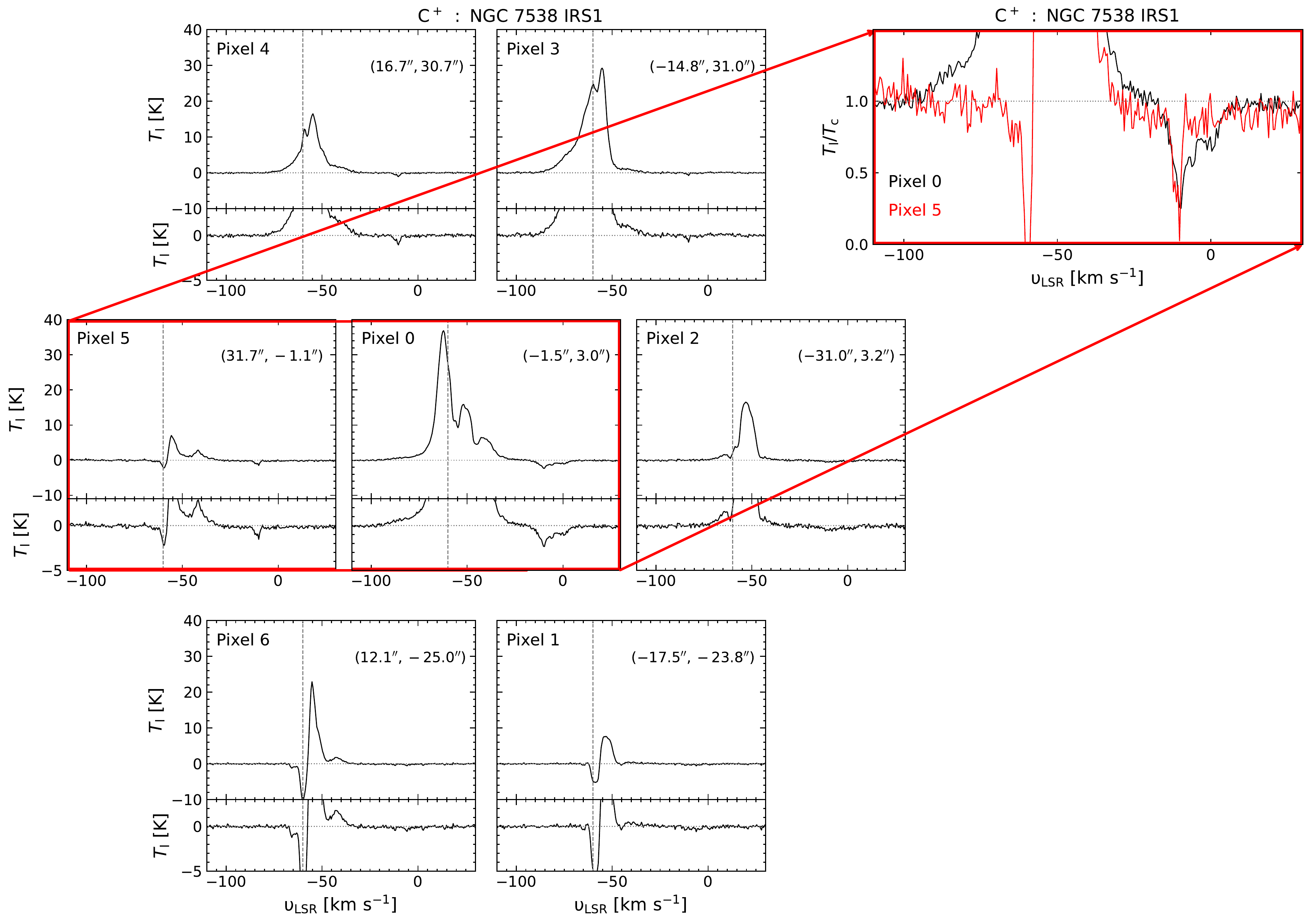}
        \caption{Left: LFA/upGREAT footprint displaying the [C\,{\small II}] spectral line profiles as observed in each of its seven pixels centered on the background continuum source NGC~7538~IRS1 in the top panels while the bottom panels zoom-in on the respective spectra to better see the absorption components. Right: Overlay comparing the line-of-sight absorption observed in the continuum normalized [C\,{\small II}] spectra obtained from pixel 0 (in black) and pixel 5 (in red).}
    \label{fig:NGC7538_CII_footprint}
\end{figure*}

Overall, variations in the column density ratios inferred from the foreground components associated with absorption against the local arm are similar for all three sources. Located at comparable distances beyond the solar circle, all three of the sight lines discussed in this paper probe similar foreground material, in particular that toward W3(OH) and W3~IRS5, both of which are star-forming regions contained in the W3 complex. While the ratios averaged over line-of-sight velocity intervals for both W3(OH) and W3~IRS5 appear consistent with each other, examining their individual spectra shows apparent variations as discussed in Sect.~\ref{subsec:spec_los_props}. A similar analysis has been carried out by \citet{Welty2020} at far-ultraviolet wavelengths in an attempt to characterize the properties and spatial structure for sight lines near HD~62542. In addition, for CH, C$^+$ and O observed using the upGREAT seven pixel array receivers, LFA and HFA, we can measure fluctuations on smaller scales for sufficiently extended sources by examining the spectral line features obtained in the outer pixels;  the latter are positioned in a hexagonal pattern around the center with a pixel spacing of $\sim 31.8^{\prime\prime}$ and ${\sim 13.6^{\prime\prime}}$ for the LFA and HFA, respectively. We illustrate this in Figure~\ref{fig:NGC7538_CII_footprint}, which displays the [C\,{\small II}] spectral line features obtained in each of the seven LFA pixels toward NGC~7538~IRS1. While we detect [C\,{\small II}] emission in each individual pixel, the continuum brightness is only sufficient in a few of the off-center pixels to detect foreground absorption components (not associated with the molecular cloud). The spectrum obtained in pixel 5, for example, shows narrow absorption features near 10~km~s$^{-1}$ associated with material in the local arm, with an absorption line strength (when normalized with respect to the continuum temperature of 1.57~K) that is comparable to that observed in the central pixel. However, with weaker continuum levels of the order of 0.1~K, we do not observe significant ($\geq3\sigma$) line-of-sight absorption in the [C\,{\small II}] spectra obtained in the other off-center pixels.

The observed fluctuations in the abundance ratios are very useful to constrain properties of interstellar turbulence: its driving scale, driving mode (solenoidal vs compressional) and its intensity \citep[e.g., the sonic Mach number discussed by][]{Bialy2017, Bialy2019}. The turbulence in cold gas is typically supersonic, thus producing significant density fluctuations in the gas, which are in turn translated into fluctuations in the molecular abundances. Thus larger dispersions in molecular abundances typically correspond to larger turbulence driving scales and higher sonic Mach numbers \citep{Bialy2017, Bialy2019, Bialy2020}. For example, in their study \citet{Bialy2019} used \textit{Herchel} observations of OH$^+$, H$_2$O$^+$ ArH$^+$ and found that toward the observed sight lines, a model of turbulence driven on large scale with superrsnoic velocities is favored \citep[see also][for an application to H and H$_2$ data]{Bialy2017, Bellomi2020}.

Furthermore, the column density ratios in turn can be translated into molecular fractions. For example toward W3~IRS5, using Eq.~\ref{eqn:molfrac} for CH, Eq~12 from \citet{Indriolo2015} for OH$^+$/H$_{2}$O$^+$ and Eq.~9 from \citet{Jacob2020Arhp} for ArH$^+$ in the foreground clouds, we find that CH traces molecular fractions between $f^N({\rm H}_{2}) \sim 0.66$ and 0.83 while OH$^+$ and ArH$^{+}$ trace gas with $f^{N}({\rm H_{2}})\sim2\times10^{-2}$ and $\sim1$-$5\times10^{-3}$, respectively, assuming an equilibrium ortho-to-para H$_{2}$O$^+$ ratio of 3. The complete HyGAL SOFIA data set will enable us to study variations in the average abundances of or molecular fractions traced by the different species studied and across different spiral arms.

\subsection{\texorpdfstring{$^{12}$}{12}C/\texorpdfstring{$^{13}$}{13}C isotopic abundances}\label{subsec:12c13c_ratio}
Owing to the lower ionization potential of atomic carbon (11.2~eV) in comparison to hydrogen (13.6~eV), ionized carbon, C$^+$, has proven invaluable as a tracer of both H{\small II} regions and cold neutral gas using its $^{2}P_{3/2}\rightarrow^{2}P_{1/2}$ fine-structure transition at 158~$\mu$m. 
As one of the key coolants in the ISM alongside the [OI] line at 63~$\mu$m, the [C\,{\small II}] 158~$\mu$m line is responsible for the transition between different ISM phases and is therefore an excellent tracer for the warm ionized medium, the diffuse atomic gas, and the diffuse molecular gas.

With a frequency separation $<1$~GHz (equivalent to a velocity separation of 130~km~s$^{-1}$, see Table~\ref{tab:13cplus}) between its hyperfine structure components, the [$^{13}$C\,{\small II}] line is detected simultaneously with [C\,{\small II}] in the 4~GHz wide band of the LFA channel on upGREAT. The [$^{13}$C\,{\small II}] signatures were first detected using the Kuiper Airborne Observatory (KAO) by \citet{Boreiko1988, Stacey1991, Boreiko1996} and then by \citet{Ossenkopf2013} using \textit{Herschel}/HIFI, for example, and more recently by \citet{Graf2012} and \citet{Guevara2020} using LFA/upGREAT. While the strongest [$^{13}$C\,{\small II}] hyperfine component corresponding to the $F=2-1$ transition blends with the broad emission wing of [C\,{\small II}], the $F=1-0$ and $F=1-1$ transitions are detected toward W3~IRS5 and NGC~7538~IRS1. 
The high spectral resolution provided by the upGREAT instrument allows us to study the $^{12}$C/$^{13}$C isotopic abundance ratio, an important diagnostic tool for probing Galactic chemical evolution or simply the nucleosynthesis history of the Galaxy. As a central species in gas phase astrochemistry, the detection of $^{13}$C$^+$ further presents an opportunity to study optical depth effects in C$^+$ as well as fractionation that affects the $^{12}$C/$^{13}$C ratio. This has important ramifications on the fractionation of subsequently formed carbon bearing species such as CH and CH$^+$.\\

\begin{figure}
    
    \includegraphics[width=0.48\textwidth]{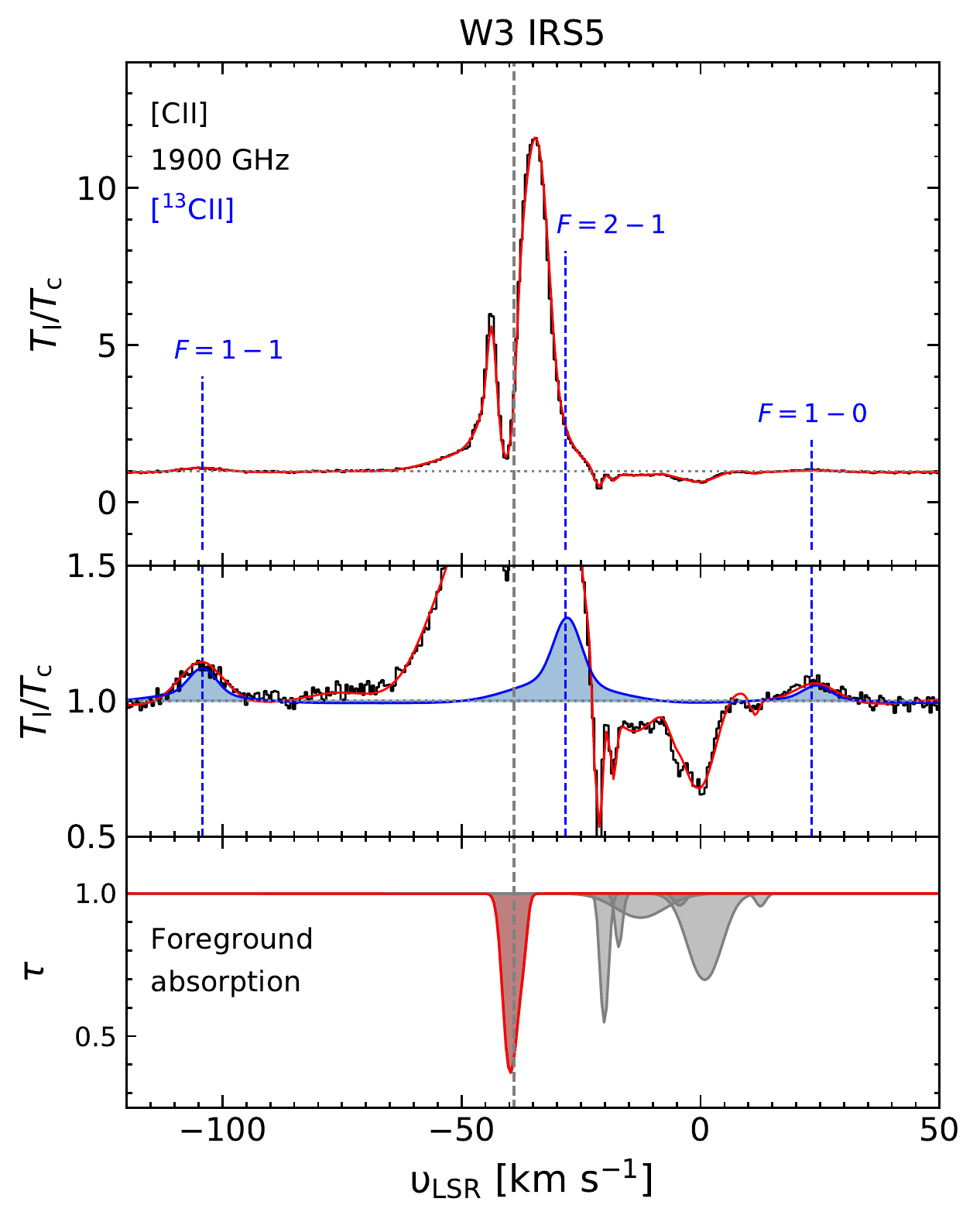}
    \caption{Demonstration of the XCLASS fits to both the [C\,{\small II}] and [$^{13}$C\,{\small II}] features toward W3~IRS5. Top: The simultaneous fit in red overlaid on the spectrum in black. Middle: A zoom-in on the spectrum and fits, where the blue shaded curves highlight the XCLASS fit to the [$^{13}$C\,{\small II}] hyperfine structure lines. Bottom: Individual [C\,{\small II}] foreground absorption components in gray with the self-absorption feature seen near the systemic velocity of the source highlighted in red.}
    \label{fig:12c13cplus_fits}
\end{figure}

\begin{figure}
    \centering
    \includegraphics[width=0.48\textwidth]{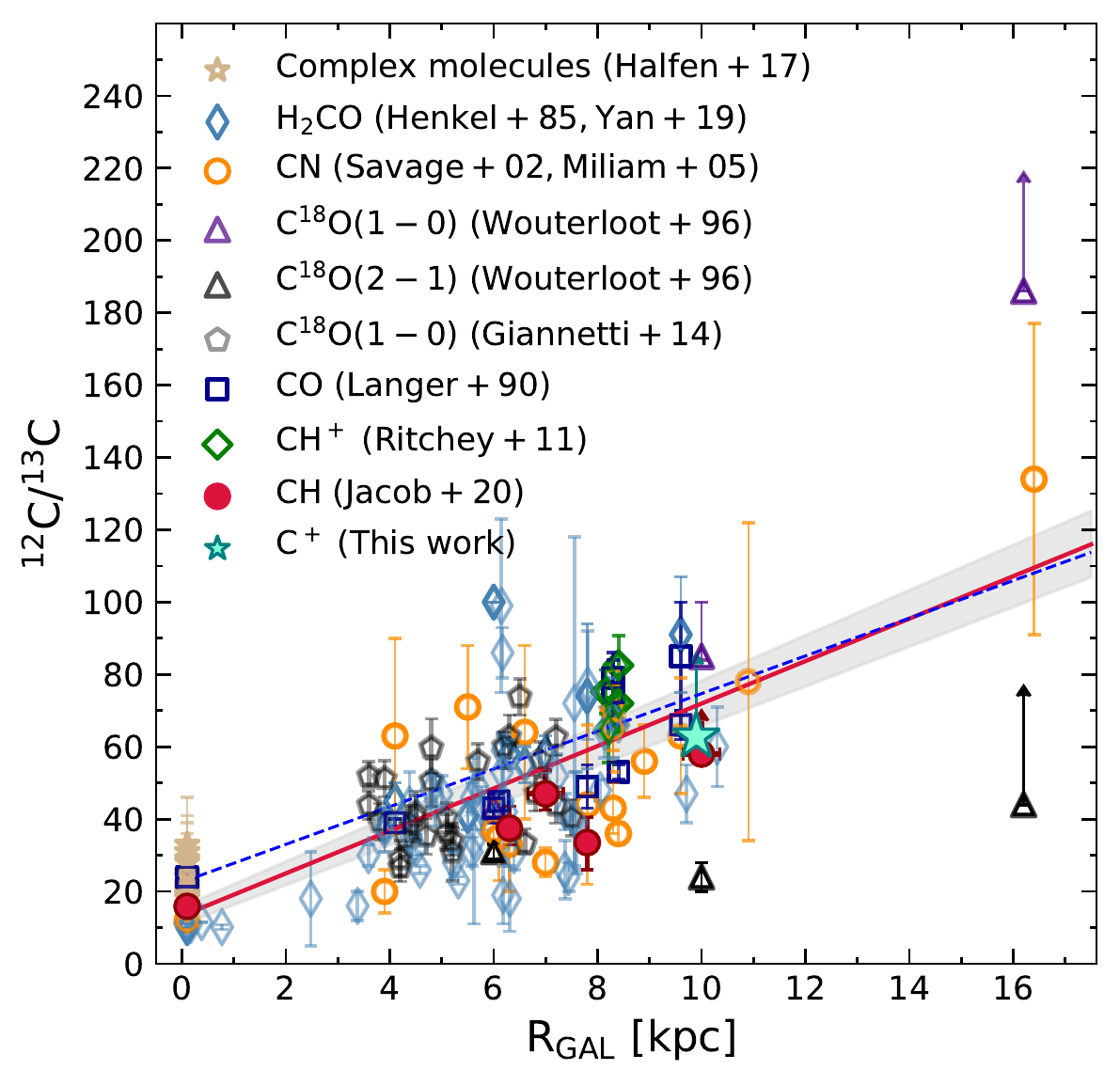}
    \caption{$^{12}$C/$^{13}$C isotopic abundance ratio as a function of Galactocentric distance, $R_\text{GAL}$~(kpc) adapted from \citet{Jacob2020}. The teal star represents the $^{12}$C/$^{13}$C ratio obtained using C$^+$ (this paper). The black solid line represents
the weighted fit to the data except that from this work with the gray shaded region demarcating the 1$\sigma$ interval of this fit. }
    \label{fig:12c13c_ratio_galacitc_gradient}
\end{figure}

Being affected by self-absorption, the [C\,{\small II}] spectra are modeled simultaneously with that of the [$^{13}$C\,{\small II}] hyperfine lines as shown in Figure~\ref{fig:12c13cplus_fits}. Using the $F=1-0$ transition as a template for the line widths and intensities (which for W3~IRS5 is fit using two components; a narrow feature ($\Delta\upsilon \sim 4.5$~km~s$^{-1}$) and a broader component at ($\Delta\upsilon \sim 11.2$~km~s$^{-1}$)), the [C\,{\small II}] line emission and self-absorption features at the velocity of the molecular cloud are modeled without using priors for the intensity of the C$^+$ emission determined from measurements of the [$^{12}$C]/[$^{13}$C] Galactic gradient. From the resulting profiles we derive a [$^{12}$C]/[$^{13}$C] ratio $>63$. Partly taking into account the optical depth effects seen in [C\,{\small II}] the [$^{12}$C]/[$^{13}$C] discussed above are higher than what they would be if the self-absorption feature was not accounted for. Figure~\ref{fig:12c13c_ratio_galacitc_gradient} compares the [$^{12}$C]/[$^{13}$C] ratios hence derived with that obtained using other species. We find the isotopic abundance ratio derived using $^{12}$C$^+$ and $^{13}$C$^+$ to lie perfectly on the Galactic gradient derived by \citet{Jacob2020}. This result is most likely fortuitous, representing only a lower limit given that the [C\,{\small II}]/[$^{13}$C\,{\small II}] is affected by fractionation effects and that there are uncertainties in the excitation temperatures used. With more detections of the [$^{13}$C{\small}] transitions toward the other targeted sources, this survey will extend our understanding of the optical depth and fractionation effects of C$^+$.  

\section{Summary} \label{sec:summary}
In this paper we introduced the SOFIA Legacy program HyGAL aimed at characterizing the Galactic ISM through absorption line studies of six key hydrides (ArH$^+$, p-H$_{2}$O$^+$, OH$^+$, SH, OH, and CH) and two atomic species C$^+$ and O toward 25 bright Galactic background continuum sources. Concerned with the lifecycle of molecular material in the universe, the HyGAL program addresses questions related to the formation of molecular clouds and investigates the processes responsible for the phase transition from atomic to molecular gas. This is achieved by making use of the well established diagnostic powers of the different targeted hydrides which range from tracers of the atomic gas and cosmic-ray ionization rates such as ArH$^+$, p-H$_{2}$O$^+$, and molecular and CO-dark H$_{2}$ gas tracers, like CH and OH, to tracers of the dissipation of turbulence such as SH.  \\

These goals are realized by taking advantage of the unique capabilities and high resolution provided by the upGREAT and 4GREAT instruments on board the SOFIA telescope. Here, we presented the observing strategy and data reduction techniques used as well as illustrated the potential analysis through the example of three sources in our sample- W3(OH), W3~IRS5 and NGC~7538~IRS1. For all three sources, the observed hydride spectra reveal widespread absorption not only at the velocities of the molecular cloud but also along its line-of-sight, except for SH likely because of the species lower abundances toward these sight lines. The spectra of the atomic species, C$^+$ and O, are more complex showing a combination of emission and absorption features toward the hot core, in addition to the line-of-sight absorption features. Despite local variations in abundances, the column densities and column density ratios with respect to both atomic and molecular hydrogen derived over specific velocity intervals that correspond to spiral-arm crossings are similar for all three sources. Extending this analysis to the complete HyGAL data set will provide column density distribution profiles for all the species observed, toward all sources studied to systematically investigate the properties of the diffuse gas sight lines over Galactic scales. The joint analysis of the SOFIA HyGAL data along with the H\,{\small I} data obtained from the JVLA, the 3~mm lines obtained from the IRAM 30~m telescope and chemical models will not only address the questions we set out to answer, but will also provide a wealth of knowledge about the background continuum sources making this survey of valuable legacy for future ISM studies. 

\begin{acknowledgments}
This work is based on observations made with the NASA/ DLR Stratospheric Observatory for Infrared Astronomy (SOFIA). SOFIA is jointly operated by the Universities Space Research Association, Inc. (USRA), under NASA contract NNA17BF53C, and the Deutsches SOFIA Institut (DSI) under DLR contract 50 OK 0901 to the University of Stuttgart.  We are immensely grateful for the outstanding support provided by the SOFIA Operations Team and the GREAT Instrument Team for this large observing effort. A.M.J., D.A.N. and M.G.W. were generously supported by USRA through a grant for SOFIA Program 08-0038. Part of this research was carried out at the Jet Propulsion Laboratory, California Institute of Technology, under a contract with the National Aeronautics and Space Administration. D.S. and S.W. acknowledges support of the Bonn-Cologne Graduate School, which is funded through the German Excellence Initiative as well as funding by the Deutsche Forschungsgemeinschaft (DFG) via the Collaborative Research Center SFB 956 ``Conditions and Impact of Star Formation" (subprojects C5 and C6). S.W. acknowledges support by the European Research Council via ERC Starting Grant no. 679852 (Radfeedback).\\
\end{acknowledgments}

\facilities{SOFIA (upGREAT, 4GREAT), \textit{Herschel} (HIFI)}

\software{GILDAS/CLASS \url{https://www.iram.fr/IRAMFR/GILDAS}, matplotlib \citep{matplotlib}, numpy \citep{numpy}, and scipy \citep{SciPy-NMeth2020}, 
         XCLASS \citep{Moller2017}.
          }

\appendix

\section{Line-of-sight properties of individual sources}\label{sec:los_source_properites}
Given that much of this work investigates the trends of various quantities with Galactocentric distance, $R_{\rm GAL}$, the spectra towards each sight line is divided into local standard of rest (LSR) velocity intervals that correspond to absorption features arising from different spiral-arm and inter-arm crossings. Galactic rotation models are used to relate the velocities of the different line-of-sight absorption components to Galactocentric distance, following the parameters determined by \citet{Reid2019}. Furthermore, for a general check on all distances, we employed the parallax-based distance calculator accessible from the website of the Bar and Spiral Structure Legacy survey (BeSSeL)\footnote{See, \url{http://www.vlbi-astrometry.org/BeSSeL/node/378})}.

\subsection*{HGAL284.015$-$00.86}
HGAL284.015$-$00.86 \citep[IRAS 10184$-$5748 or Dutra-45,][]{Dutra2003} is a very small irregularly shaped ultra-compact H{\small II} region (as per the classification by \citet{Wood1989a}) located on the western edge of an optical nebula at a distance of 5.7~kpc on the Carina arm. Extensive observations of the CS $J=2-1$ transition \citep{Bronfman1996} toward this region reveals HGAL284.015$-$00.86 to be a promising region which can be used to investigate mass segregation effects at the birth sites of massive stars. With only a handful of molecular line surveys being carried out toward this region, the velocity range covered by the line-of-sight absorption is unknown.

\subsection*{HGAL285.26$-$00.05}
HGAL285.26$-$00.05 is an embedded infrared cluster 1$\rlap{.}^{\prime}$5 wide, commonly referred to as Dutra, Bica, Soares, Barbuy 48 - DBSB 48 \citep{Dutra2003}, located in the Hoffleit 18 nebula discovered by \citet{Hoffleit1953}. Prior to its detection at infrared wavelengths using the Two Micron All Sky Survey (2MASS), this nebula was identified as an H{\small II} region using radio observations by \citet{Caswell1987}. Located roughly at the center of the Sagittarius–Carina arm at a distance of 4.3~kpc in the fourth quadrant of the Galaxy, the line-of-sight toward this source does not cross any other spiral arm. 
\subsection*{G291.579$-$00.431}
NGC~3603 is a massive OB cluster and one of the most luminous H{\small II} regions in the Galaxy (100 times more luminous than the Trapezium cluster in Orion; \citet{Goss1969}). Associated with NGC~3603, G291.579$-$00.431 is located in the Sagittarius-Carina spiral arm at a distance of approximately 8~kpc \citep{Lee2012}, corresponding to one the strongest masing sources of the region. Lying tangential to the Carina arm this source only covers a short sight line with absorption features typically observed between $-2$ and $23$~km~s$^{-1}$.

\subsection*{IRAS~12326$-$6245}
The infrared source IRAS 12326$-$6245, with a $\upsilon_{\rm LSR} \sim -40$~km~s$^{-1}$, located tangential to the Scutum-Centaurus arm, located 7.2~kpc from the Galactic center \citep{Green2011}. Deeply embedded in a dense molecular cloud, IRAS 12326$-$6245 harbors in its center one of the most energetic and massive bipolar molecular outflows known among objects of similar luminosity \citep{Henning2000}. The line-of-sight crosses the Sagittarius-Carina arm near $-22$~km~s$^{-1}$.

\subsection*{G327.3$-$00.60}
Located in the Scutum-Centaurus-Crux arm, the G327.3$-$00.60 star-forming region is at a distance of 3.1~kpc \citep{Wienen2015} and hosts different evolutionary phases of massive star formation  consisting of a bright H{\small II} region and a chemically rich hot core in dust condensates within a $\sim 3$~pc region \citep{Wyrowski2006}. The sight line toward this target shows absorption against the molecular cloud at a $\upsilon_{\rm LSR} = -47$~km~s$^{-1}$ as well as absorption near $-$10~km~s$^{-1}$ coinciding with the crossing of the Sagittarius-Carina arm.

\subsection*{G328.307+00.423}
The G328.307+00.423 cluster harbors massive young stellar objects that are highly concentrated within the central region, with intermediate-mass YSOs more spread out. Situated in the fourth quadrant, G328.307+00.423 is a bright ATLASGAL \citep{Schuller2009} $870~\mu$m source at a distance of 5.8~kpc \citep{Urquhart2018} and located in the Norma spiral arm. The sight line towards this source contains contributions from the molecular cloud and the near side of the Norma arm between $-110$ and $-75$~km~s$^{-1}$, the Scutum-Centaurus arm between $-62$ and $-50$~km~s$^{-1}$ and the Sagittarius arm between $-45$ and $-35$~km~s$^{-1}$. In addition to these features, certain molecular species show a narrow absorption feature centered at $-15$~km~s$^{-1}$, likely associated with local gas. 

\subsection*{IRAS~16060$-$5146}
Located at a distance of 5.3~kpc \citep{Urquhart2018}, IRAS~16060$-$5146 (G330.954$-$00.182) is part of the G331.394$-$00.125 complex. It is a luminous starforming region that lies close to G328.307+00.423 in the Norma spiral arm. High angular resolution continuum observations towards this region shows the presence of a compact H{\small II} region with a complex and irregular morphology \citep{Walsh1998}. With its peak position coincident with a dust core, the H{\small II} region is extremely embedded and therefore undetectable at near-infrared wavelengths. It has an absorption profile that extends from $-130$ to +20~km~s$^{-1}$. The gas between $-60$ and +20~km~s$^{-1}$ traces the near and far-side crossings of both the Scutum-Centaurus as well as the Sagittarius spiral arms. 

\subsection*{IRAS~16164$-$5046}
The sight line towards IRAS~16164$-$5046 (G332.826$-$00.549), which has a near kinematic distance of 3.6~kpc \citep{Moises2011}, primarily traces the Scutum-Centaurus and Sagittarius spiral arms. The strongest absorption features are near $-40$~km~s$^{-1}$ and are often observed alongside redshifted absorption that traces infalling material, similar to that seen in other sources like G19.61$-$0.23. 

\subsection*{IRAS~16352$-$4721}
At a distance of ${\sim}12.3$~kpc \citep{Green2011}, IRAS~16352$-$4721 (G337.704$-$00.054) is an embedded young-stellar object, and one of the most distant Galactic sources in our sample. Absorption features cover a range of velocities from $-135$ to ${\sim}$ +20~km~s$^{-1}$. The line-of-sight  absorption at negative velocities between $-$140 and $-$60~km~s$^{-1}$ arises mainly from the 3~kpc and the Norma spiral arms with relatively weaker contributions from the other arms as well. 

\subsection*{IRAS~16547$-$4247}
IRAS~16547$-$4257 (G343.12$-$0.06) at 2.7~kpc \citep{Giannetti2014} on the Scutum arm has the distinction of being the first reported, and most luminous young stellar object known to harbor a radio jet. The jet, located at the center of the molecular core, drives a highly energetic and collimated bipolar outflow over 1.5~pc \citep{Garay2003}. In addition to the features associated with the molecular cloud, itself at $\upsilon_{\rm LSR}$ between $-50$ and $-30$~km~s$^{-1}$, the line-of-sight toward this source is also known to show absorption from $-25$ to $-15$~km~s$^{-1}$ associated with the Sagittarius arm and the local arm near 0~km~s$^{-1}$.

\subsection*{NGC~6334~I}
NGC~6334 is a filamentary, star-forming cloud on the Sagittarius spiral arm at a distance of 1.35~kpc \citep{Immer2013}. It harbors a string of dense cores at far-infrared wavelengths that are identified using roman numerals from I to VI \citep{McBreen1979}. Source I (also designated as NGC~6334~F) is a prototypical hot molecular core at the head of a cometary ultra-compact H{\small II} region with a rich molecular inventory. An additional source was identified $\sim 2^{\prime}$ north of source I, named `I(N)' \citep{Gezari1982G}. While both cores are sites of active star-formation, NGC~6334~I is in a relatively more evolved stage of star formation than NGC~6334~I(N). Absorption components observed in the velocity range from $-17$ to 3~km~s$^{-1}$ can be ascribed to the absorption against the core itself and foreground components associated with the NGC~6334 molecular cloud. In addition, this source's line-of-sight covers foreground absorption arising from two unrelated clouds between 5 and 10~km~s$^{-1}$ \citep{vdWiel2016}.

\subsection*{G357.558$-$00.321}
 G357.558$-$00.321 is an H{\small II} region located in the vicinity of an unusual supernova remnant, the `Tornado nebula', G357.7$-$0.1 (MSH 17$-$39) in the central 1~kpc of the Galaxy \citep{Frail1996}. Spatially, G357.7$-$0.1 consists of a bright head and a faint tail, as well as a compact source 30$^{\prime\prime}$ north of the head (known as the eye and not associated with the remnant \citep{Sakai2014}). Owing to difficulties in determining meaningful kinematic distances toward this direction using H\,{\small I} absorption measurements, the distance adopted toward this region has been determined from a single maser measurement near $\upsilon_{\rm LSR}$ = -12.4~km~s$^{-1}$, detected along the western edge of MSH 17$-$39, just north of the compact source. However, very little is known about G357.558$-$00.321 itself.

\subsection*{HGAL0.55$-$0.85}
The HGAL0.55$-$0.85 molecular cloud region is located in the Galactic center region \citep{Walsh1998}. A site of active star-formation, this region hosts two Class II methanol masers separated by $\sim 3^{\prime\prime}$ with associated NH$_3$ emission \citep{Hill2005, Beuther2009}. However, as it has been the target of only a handful of astronomical studies, very little else is known about this region.

\subsection*{G09.62+0.19}
The G09.62+0.19 complex, at a distance of 5.2~kpc \citep{Sanna2009}, is located in the Norma arm close to the expanding 3~kpc arm and contains a cluster of radio continuum sources, denoted from A to I. Since each of these continuum sources is in a different evolutionary stage, the overall morphology of the complex is suggestive of sequential high-mass star-formation progressing from the most evolved H{\small II} region, component A -- hosting massive star-formation -- to regions like component E that show signs of a very early phase of massive star formation \citep{Hofner1994, Liu2017}. Beyond the systemic velocity of the source $\upsilon_{\rm LSR} = 4.3~$km~s$^{-1}$, the source also shows weaker line-of-sight absorption against the Scutum-Centaurus and Sagittarius spiral arms between 10 and 40~km~s$^{-1}$.

\subsection*{G10.47+0.03}

The G10.47+0.03 H{\small II} region contains three ultra-compact H{\small II} regions, G10.47+0.03 A, B and C, and is located at a distance of 8.55~kpc \citep{Sanna2014}. The sight line towards G10.47+0.03 therefore crosses several spiral arms in the inner Galaxy, resulting in a broad absorption spectrum covering a $\upsilon_{\rm LSR}$ range from $-$35 to 180~km~s$^{-1}$. Absorption at $\upsilon_{\text{LSR}}$ between $-35$ and $48$~km~s$^{-1}$ arises partly from inter-arm gas and partly from the near-side crossing of the Sagittarius and Scutum-Centaurus arms. Features at $\upsilon_{\text{LSR}} > 80~$km~s~$^{-1}$ are associated with the 3~kpc arm and the Galactic bar with contributions from the Norma arm at intermediate velocities (40~km~s~$^{-1} < \upsilon_{\text{LSR}} < 80~$km~s~$^{-1}$). Higher velocity components at $\upsilon_{\text{LSR}} > 120~$km~s~$^{-1}$ likely trace gas that lies beyond the Galactic center belonging to the 135~km~s$^{-1}$ arm \citep{Sormani2015}. 
 
\subsection*{G19.61$-$0.23}
G19.61$-$0.23 is a high-mass star-forming region, known to harbor several ultra-compact H{\small II} regions and a hot molecular cloud \citep{Furuya2011}. Rich in young stellar objects, G19.61$-$0.23 has been the target of several studies aimed at understanding cluster formation and other processes at the early phases of the high-mass star-formation. This massive star-forming clump is at a heliocentric distance of 12.7~kpc \citep{Urquhart2014} with an absorption spectrum extending from $-10$ to 150~km~s$^{-1}$. The absorption at $\upsilon_{\rm LSR}$ between $-10$ and 20~km~s$^{-1}$ primarily traces inter-arm gas whereas absorption dips between 35 and 74~km~s$^{-1}$ arise from the near- and far-side crossings of the Sagittarius spiral arm, where the systemic velocity of the source is at 40~km~s$^{-1}$. Signatures of infalling gas have previously been studied by \citet{Furuya2011}, whose observations of the $J=3\rightarrow2$ transitions of $^{13}$CO and $^{18}$CO show inverse P-Cygni profiles or red-shifted absorption alongside a blue-shifted emission component. Narrow absorption features at 80 and 113~km~s$^{-1}$ possibly arise from the edge of the Sagittarius-Carina and Scutum-Centaurus spiral arms, respectively.

\subsection*{G29.96$-$0.02}
One of the brightest submillimeter continuum sources in the Milky Way galaxy, G29.96-0.02 is a classical example of a compact H{\small II} region with a cometary-like morphology \citep{Wood1989} caused by its interaction with a molecular core ${\sim 2\rlap{.}^{\prime\prime}6}$ away. While the main exciting source of the compact H{\small II} region has been identified in the near-infrared as an O5-O6 star \citep{Watson1997}, additional sources have been detected toward the rim of the H{\small II} region, indicating an embedded cluster \citep{Pratap1999}. The rich molecular inventory of this region has not only been revealed through the characterization of strong emission lines but also through absorption line measurements of species such as HF \citep{Kirk2010} and ortho-H$_{2}$Cl$^+$. At a distance of 6.7~kpc \citep{Urquhart2018}, G29.96$-$0.02 lies along the Galactic bar and on the Norma arm. In addition to absorption at molecular cloud velocities between 87 and 120~km~s$^{-1}$, the sight line velocity distribution also includes absorption arising from the Sagittarius near- and far-side between 37 and 85~km~s$^{-1}$, and the Aquila Rift near 9~km~s$^{-1}$.

\subsection*{G31.41+0.31}
A prototypical hot molecular core, G31.41+0.31 is located on the Scutum-Centaurus arm at a distance of 5~kpc \citep{Moises2011}, with the hot core separated by $\sim 5^{\prime\prime}$ from an ultra-compact H{\small II} region. The high bolometric luminosity ($\approx 3\times10^{5}~L_{\odot}$, \citet{Cesaroni2011} and references therein) and extended halo hints toward the presence of a disk-like structure rotating about young-stellar objects. The presence of such a rotating toroid, together with the possibility of outflows from the center of the molecular core complicates the analysis of spectral line features toward this cloud component at $\upsilon_{\rm LSR} = 98~$km~s$^{-1}$, often distorting the observed line profiles or showing self-absorption \citep{Colzi2017}. The line-of-sight toward G31.41+0.31 crosses the Sagittarius arm at velocities roughly between 30 and 55~km~s$^{-1}$ and the local arm at velocities $<20$~km~s$^{-1}$. 

\subsection*{W43~MM1}
W43~MM1 is a high mass proto-stellar object located at a distance of 5.5~kpc \citep{Zhang2014} near the end of the Galactic bar, in the mini-starburst region of W43. While it resembles a young protostar in its main accretion phase, W43~MM1 has already developed a hot core with temperatures higher than 200~K. In spite of its many peculiarities, several absorption studies have been previously carried out toward this source; for example, it was found to be rich in water and its isotopologs in the WISH program \citep{vDishoeck2011, Herpin2012}. 
Located in a crowded part of the Galactic plane, molecular line data toward this source exhibits absorption features between 0~km~s$^{-1}$ and 140~km~s$^{-1}$ with extended absorption near $\upsilon_{\rm LSR} = 98~$km~s$^{-1}$ associated with absorption against the continuum background source and narrower features arising from line-of-sight components at $\upsilon_{\rm LSR}< 85~$km~s$^{-1}$. This includes components near 40~km~s$^{-1}$, 10~km~s$^{-1}$ and 0~km~s$^{-1}$ corresponding to the near-side crossing of the Sagittarius arm, the Aquila Rift, and the local arm, respectively. 

\subsection*{G32.80+0.19}
The G32.80+0.91 H{\small II} region has a complex structure with at least three associated components; a bright compact source labeled A, an extended component with a cometary-like morphology located 6$^{\prime\prime}$ northeast of A called B; and a weaker component to the northwest which is unresolved in continuum maps at 8.3~GHz, dubbed C \citep{Garray1994, Gomez1995}. Situated in the Perseus arm 13~kpc away, the sight line toward this H{\small II} region consists of several spiral arm components spanning a range of velocities from 0 to 110~km~s$^{-1}$. Beyond the systemic velocity of the source at ${\sim 15}$~km~s$^{-1}$, the line-of-sight crosses the near- and far-side of the Sagittarius arm between 25 and 65~km~s$^{-1}$, showing narrow absorption at 71~km~s$^{-1}$ corresponding to absorption against the Aquila rift and finally the Scutum-Centaurus arm at $\upsilon_{\rm LSR}>$78~km~s$^{-1}$. 

\subsection*{G45.07+0.13}
At a distance of 4.3~kpc, G45.07+0.13 is a pair of spherical ultra-compact H{\small II} regions showing OH, H$_2$O, and CH$_3$OH maser emission located in the Sagittarius-Carina spiral-arm with at least three continuum sources observed in the mid-infrared \citep{deBuizer2005}. Furthermore, observational evidence distinguishes the G45.07+0.13 core from neighboring cores such as G45.12+0.13 as a relatively young site of massive star formation \citep{Hunter1997}. In addition to absorption against the background continuum source at $\upsilon_{\rm LSR} = 60~$km~s$^{-1}$ and the local arm near 0~km~s$^{-1}$, some species show narrow absorption components between $\upsilon_{\rm LSR} =20$ and 60~km~s$^{-1}$  tangentially along the Sagittarius arm.

\subsection*{DR21}
Located in the Cygnus X complex at a distance of 1.5~kpc \citep{Rygl2012}, the DR21 molecular ridge is a dense region of massive star-formation at $\upsilon_{\rm LSR}=-3$~km~s$^{-1}$. This star-forming region comprises of a group of several compact H{\small II} regions and is known to host highly energetic molecular outflows. This outflow is likely driven by an atomic jet as evidenced by the detection of high velocity H\,{\small I} emission (up to 90~km~s$^{-1}$) using the VLA \citep{Russell1992}. In addition to the molecular cloud associated with the DR21 region itself, the filament contains another center of activity corresponding to the OH maser source, DR21(OH), which lies about 3$^{\prime}$ (or a projected distance of 1.3~pc) north of DR21. Molecular line observations reveal a prominent blue-shifted outflow wing at velocities $<-17$~km~s$^{-1}$ while the corresponding red-shifted wing is more difficult to discern \citep{Schneider2010}. Furthermore, in addition to the foreground absorption arising from the Cygnus X complex, there is a foreground component at $\upsilon_{\rm LSR}\sim 9~$km~s$^{-1}$ associated with W75~N star-forming region located northwest of the filament.

\section{Spectroscopic parameters of \texorpdfstring{$^{13}{\rm C}^+$}{13C+}}\label{sec:13cplus}
\begin{table}[ht]
    \caption{Spectroscopic parameters of the \texorpdfstring{$^{13}{\rm C}^+$}{13C+} transition near 158~$\mu$m.}
    \begin{center}
    \begin{tabular}{ll cl cl}
    \hline\hline
         Transition & Frequency & $A_{\rm u,l}$ & $E_{\rm u}$ & Velocity offset\tablenotemark{a} & Relative\\
         $F^{\prime}-F^{\prime\prime}$ & [GHz] & \multicolumn{1}{c}{$10^{-6}\times$[s$^{-1}$]} & [K] & \multicolumn{1}{c}{[km~s~$^{-1}$]} & Intensity\\
         \hline
         1-1 & 1900.136(10) & 0.77 & 91.2& +63.2 & 0.125\\
         2-1 & 1900.466(2) & 2.32& 91.2 & +11.2 & 0.625 \\
         1-0 & 1900.950(15) & 1.55 & 91.2& $-65.2$ & 0.250 \\

         \hline
    \end{tabular}
   
    \tablenotetext{a}{The velocity offsets are computed with respect to the [C\,{\small II}] line at 1900.536~GHz.}
     \end{center}
    \label{tab:13cplus}
\end{table}

\bibliography{ref}{}
\bibliographystyle{aasjournal}

\end{document}